\documentclass[aps,prb,reprint,superscriptaddress]{revtex4-2}
\usepackage{hyperref}
\hypersetup{
    colorlinks = true,
    linkcolor = [rgb]{0.70, 0.13, 0.13},
    citecolor = [rgb]{0.25, 0.41, 0.88},
    urlcolor = [rgb]{0.25, 0.41, 0.88}
}
\usepackage{amsmath}
\usepackage{amsfonts}
\usepackage{amssymb}
\usepackage{pifont}
\usepackage{mathtools}
\usepackage{bm}
\usepackage{cases}
\usepackage{braket}
\usepackage[version=4]{mhchem}
\usepackage{graphicx}
\allowdisplaybreaks%

\graphicspath{{figures/}}

\DeclareMathOperator{\Tr}{Tr}

\begin{document}
\title{Nematic correlations and nematic Berezinskii-Kosterlitz-Thouless transition \\
       in spin-1 kagome lattice antiferromagnets}

\author{Chun-Jiong Huang}
\thanks{These authors contributed equally.}
\affiliation{Department of Physics and HKU-UCAS Joint Institute for Theoretical and Computational Physics at Hong Kong, 
The University of Hong Kong, Hong Kong, China} 
\affiliation{The University of Hong Kong Shenzhen Institute of Research and Innovation, Shenzhen 518057, China}
  
\author{Xu-Ping Yao}
\thanks{These authors contributed equally.}
\affiliation{Kavli Institute for Theoretical Sciences, University of Chinese Academy of Sciences, Beijing 100190, China}

\author{Gang v.~Chen}
\email{chenxray@pku.edu.cn}
\affiliation{International Center for Quantum Materials, School of Physics, Peking University, Beijing 100871, China}
\affiliation{Collaborative Innovation Center of Quantum Matter, Beijing 100871, China}

\date{\today}

\begin{abstract}
    Nematicity plays an important role in strongly correlated electron systems. 
    We explore the spin nematicity of a spin-1 kagome lattice antiferromagnet with 
    the bilinear-biquadratic model and single-ion anisotropy using a generalized semiclassical  
    approximation and Monte Carlo simulations. We reveal a rich ground state phase diagram, 
    characterized by two main regions: 
    a pure spin nematic phase and a region featuring the coexistence of a classical spin liquid 
    and ferroicities for both dipolar and quadrupolar moments. 
    The thermal fluctuation melts the spin nematic order into a critical phase 
    with a quasi-long-range nematic order.
    Due to the fluctuating vortices of the spin nematic order, 
    this critical phase further undergoes a nematic Berezinskii-Kosterlitz-Thouless transition to a paramagnetic phase, 
    marked by an anomalous stiffness jump. 
    Additionally, the single-ion anisotropy leads to weak ferromagnetism, 
    resulting in spontaneous time-reversal symmetry breaking at very low temperatures.        
    Remarkably, both two types of ferroic ordering are accompanied by classical spin liquid behaviors.
    Our results provide an intriguing glimpse into the interplay between geometric frustration 
    and intertwining spin orders with different ranks, and are 
    expected to stimulate further studies on spin-1 systems and relevant materials.
\end{abstract}
    
\maketitle

\section{Introduction}

Nematic liquid crystals are a well-known subject 
in modern soft condensed matter physics~\cite{ANDRIENKO2018520}. Their emergence and fluctuations 
in strongly correlated electronic systems have advanced 
our understanding of high-temperature superconductivity and fermion criticality. 
Electronic nematicity is often associated with the orientation of fermion bilinears 
and thus breaks spatial symmetry~\cite{Fradkin_2010}. 
It often does not directly impact electron spins, except through spin-orbit coupling. 
Although spin nematicity could occur spatially, 
it could emerge by breaking the internal spin rotational 
symmetry~\cite{Zhitomirsky_2010,PhysRevB.82.174440,PhysRevB.84.094420,PhysRevB.88.184430,PhysRevB.107.L140403,PhysRevLett.130.116701}. 
In this work, we address the fluctuations and correlations of the spin nematicity 
of a spin-1 kagome lattice antiferromagnet.

Kagome systems~\cite{10.1143/ptp/6.3.306} have been an enduring focus
in quantum materials due to their versatility in realizing novel quantum phases. 
With such a unique geometry, the electronic structure features many nontrivial ingredients,  
including flat bands, Dirac band touching, and Van Hove singularities.  
Once the correlation effects are considered, 
the intertwining and competition between these ingredients 
trigger the emergence of a plethora of exotic phenomena, 
including electron band topologies~\cite{doi:10.1126/science.aav2873,doi:10.1126/science.aav2334,Ye2018,Yin2020}, 
the quantum anomalous Hall effect~\cite{PhysRevLett.115.186802,Liu2018}, 
the fractional quantum Hall effect~\cite{PhysRevLett.106.236802}, 
unconventional superconductivity and density wave 
orders~\cite{PhysRevB.79.214502,PhysRevLett.125.247002,Jiang2021,PhysRevX.11.031050,Chen2021}, 
and so on~\cite{Yin2022}. On top of these, the geometric frustration further complicates 
the kagome physics and provides a promising route toward the long-sought quantum 
spin liquids~\cite{PhysRevB.45.12377,Balents2010,doi:10.1126/science.aay0668}. 
Major efforts have been devoted to the spin-1/2 kagome lattice Heisenberg antiferromagnet, 
and a conclusion about the quantum spin liquid has not yet been reached. 
In contrast to the spin-1/2 moment, the spin-1 moment has a larger local Hilbert space and 
allows for more possibilities for new physics~\cite{PhysRevLett.59.799,PhysRevLett.109.016402,PhysRevResearch.2.033260,PhysRevB.98.045109,PhysRevB.96.020412,PhysRevB.100.045103,PhysRevLett.120.057201,PhysRevB.100.140408,PhysRevB.102.121102,PhysRevB.105.L060403,PhysRevB.106.195147}. There have been several spin-1 kagome lattice 
antiferromagnets. Although the candidate materials Na$_2$Ta$_3$Cl$_8$~\cite{PhysRevLett.124.167203}, $m$-MPYNN$\cdot$BF$_4$~\cite{doi:10.1143/JPSJ.66.961,doi:10.1143/JPSJ.73.796,doi:10.1143/JPSJ.79.093701},
and KV$_3$Ge$_2$O$_9$~\cite{doi:10.1143/JPSJ.81.073707,PhysRevB.95.104416} undergo a structural transition and/or lattice distortion
at low temperatures and thus favor the trimerized magnetic phase due to the spin-lattice coupling~\cite{PhysRevLett.124.167203}, the newly synthesized compound $\beta$-{BaNi$_3$(VO$_4$)$_2$(OH)$_2$}
seems to retain perfect kagome lattice geometry~\cite{Li_2023}. 
Experimental study of them is quite limited at this stage, 
and more feedback on the theoretical understanding is needed.

We consider the Heisenberg model with a biquadratic interaction and single-ion anisotropy 
for a spin-1 kagome lattice antiferromagnet. This model is the simplest and most generic model 
for spin-1 moments~\cite{doi:10.1126/science.1114727,doi:10.1143/JPSJ.75.083701, 
PhysRevLett.97.087205,PhysRevB.79.214436,Kartsev2020,PhysRevB.107.L140403,PhysRevResearch.6.033077,pohle2025abundancespinliquidss1}, 
and the biquadratic interaction provides the direct interaction for the spin nematics. 
The antiferromagnetic bilinear and ferromagnetic biquadratic interactions play 
quite different roles in frustrated magnets. The former exhibits 
an extensively degenerate ground state in the classical limit. 
This classical degeneracy is characterized by the vanishing spin  
order within each triangular plaquette, accompanied by a finite configurational 
entropy density and a power-law decay of correlations. 
This physics is often called classical spin liquids (CSLs)~\cite{PhysRevLett.80.2929,PhysRevB.58.12049,annurev-conmatphys-070909-104138}. 
In contrast, the ferromagnetic biquadratic interaction simply favors a ferroquadrupolar (FQ) order. 
This spin nematic type of order breaks the continuous spin symmetry
and is forbidden 
at finite temperatures by the Hohenberg-Mermin-Wagner theorem~\cite{Mermin1966absence,Hohenberg1967existence}. 
Instead, a Berezinskii-Kosterlitz-Thouless (BKT) transition 
from the quasi-long-range ferroquadrupolar (spin nematic) order is expected just as 
an analogy to the well-understood classical XY 
model~\cite{Kosterlitz1973ordering,Kosterlitz1974the,Amit1980renormalisation}. 
The low-energy description of the spin-nematic ferroquadrupolar state  
is an unconventional nonlinear $\sigma$ model of the degenerate space 
of the real projective plane ${\mathbb{R}P^2 \simeq S^2/\mathbb{Z}_2}$~\cite{PhysRevB.68.052401,PhysRevLett.100.047203}. 
In this target space, the topological charge is defined by the first homotopy group 
${\pi_1(\mathbb{R}P^2) = \mathbb{Z}_2}$. The BKT behavior is driven 
by the proliferation of these $\mathbb{Z}_2$ half vortices with a fractional 
vorticity of $1/2$ instead of 1 in the XY case. 

In this work, we show that the CSL and BKT physics concurrently 
and naturally arise in the spin-1 model on a kagome lattice 
and reveal their intricate intertwining. 
At the zero temperature and for a pure FQ state with dominant $K/J$, the analysis of quantum fluctuations indicates an instability at $K_c/J=2$, where a flat band of spin excitations reaches zero energy and manifests the CSL correlation in the dipole channel while the FQ correlation persists. 
After the phase transition, the nonzero dipolar components define three different phases at the semiclassical level in the presence of single-ion anisotropy. 
For all three phases, there is a nontrivial extensive degeneracy on the kagome lattice. 
Specifically, for both dipoles and quadruples, a ferromagnetic (ferroquadrupolar) order can be separated, leaving a degenerate part whose summation is zero within any triangular plaquette. 
This unusual coexistence is confirmed by our semiclassical Monte Claro simulations. 
We further establish the finite-temperature phase diagram and investigate the evolution of coexisting states versus temperature.  
An anomalous BKT transition driven by the half vortices is numerically advocated in both pure FQ and coexistence regimes for $K/J \gtrsim 0.546$. 
Moreover, the CSL behavior is carefully studied at low temperatures. 
After a conventional crossover from the paramagnetic phase, pinch-point singularities in the dipole and quadruple spin correlations are developed and maintained to the lowest temperature in the simulation. 
Although there is an Ising transition due to the single-ion anisotropy at lower temperatures which is accompanied by spin-correlation enhancements at the $\bm{\Gamma}$ and $\mathbf{K}$ points, a thermal order-by-disorder effect for CSLs cannot be determined.

The rest of this work is organized as follows. 
Section.~\ref{sec:model_and_gs} describes the bilinear-biquadratic model for the spin-1 moments on the kagome lattice with an easy-axis single-ion anisotropy. 
Then, SU(3) spin operators are introduced to represent the spin-1 model in a more convenient form. 
Based on this representation, the SU(3) coherent states are defined and treated by a semiclassical  approximation. 
With the semiclassical energy optimization, the ground-state phase diagrams are obtained and presented in Fig.~\ref{fig:phasediagram_gs}(a). 
The ground-state degeneracy and coexistence are also discussed in detail. 
In Sec.~\ref{sec:finite_temperature}, a semiclassical Monte Carlo method is introduced to simulate the finite-temperature behavior. 
A representative phase diagram of spin exchange interactions and temperature is shown in Fig.~\ref{fig:phasediagram_gs}(b). 
The various phase transitions, especially the anomalous BKT transition and the Ising transition, are thoroughly studied. 
An analysis of the possible thermal order-by-order effect is also carried out numerically. 
Finally, we discuss the fruitful physics of this model and future research directions in Sec.~\ref{sec:discussion}.

\section{Bilinear-biquadratic model and semiclassical ground states}%
\label{sec:model_and_gs}

\subsection{Model and semiclassical approximation}%
\label{subsec:model}

The bilinear-biquadratic (BBQ) model for a spin-1 kagome lattice 
antiferromagnet with easy-axis single-ion anisotropy is given as 
\begin{equation}
\label{eq:hamiltonian}
    \mathcal{H} = \sum_{\braket{ij}} \left[ J \,  \mathbf{S}_i \cdot \mathbf{S}_j 
               -  K (\mathbf{S}_i \cdot \mathbf{S}_j)^2 \right]
               -  \sum_{i} D_z (S_i^z)^2.
\end{equation}
Physically, the biquadratic interaction could arise from the high-order perturbation of the Hubbard model, 
or effectively from the spin-lattice coupling by integrating out the bond phonon 
modes~\cite{doi:10.7566/JPSJ.87.023702,PhysRevB.101.024418,PhysRevLett.124.167203}.
It has a strong influence on the low-dimensional weak Mott insulators, 
particularly the two-dimensional van der Waals magnets~\cite{Kartsev2020}. 
Often, it is subordinate to the Heisenberg interaction in strength, 
but is important for stabilizing the spin nematicity. In addition,   
ultracold atomic gases or polar molecules on optical lattices  
provide an ideal platform to reach high internal spin degrees of freedom (DOFs)  
and thus allow for the exploration of a general BBQ model 
beyond the perturbation regime~\cite{PhysRevB.87.081106}. 
The single-ion anisotropy arises from the planar geometry of the magnets,
and can also be realized and tuned with ultracold atoms~\cite{PhysRevLett.126.163203}. 
 

\begin{figure*}[t]
    \centering
    \includegraphics[width=1.0\linewidth]{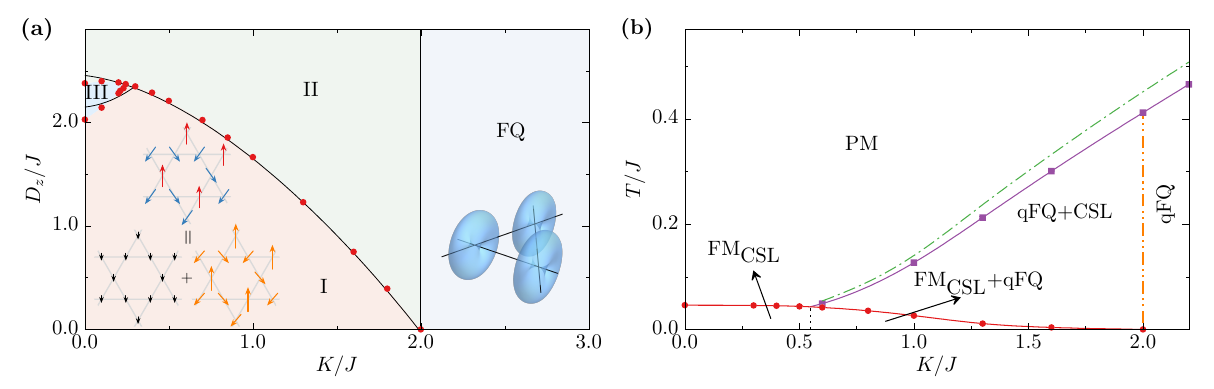}
    \caption[]{(a) Ground-state phase diagram with boundaries determined by energy optimization (lines) and simulated annealing (dots) using the semiclassical Monte Carlo method.
    The inset in the ferroquadrupolar (FQ) phase indicates the probabilities $|\braket{\bm{\Omega}|\bm{Z}}|^2$ on a triangular plaquette. 
    Explicit spin configurations for each phase and the definition of the SU(3) coherent state $\ket{\bm{\Omega}}$ can be found in the SM~\cite{SM}. 
    The decomposition of dipolar moments $\braket{\mathbf{S}_i}$ for phase I is shown in the left inset. 
    A uniform ferromagnetic (FM) component can be singled out so that the rest sum to zero on each triangle plaquette.
    (b) Finite-temperature phase diagram with ${D_z/J = 1.0}$. 
    There are four phases: the paramagnetic (PM) phase, the quasi-long-range FQ (qFQ) + classical spin liquid (CSL) phase, the FM$_\text{CSL}$+qFQ phase and the FM$_\text{CSL}$ phase for weak $K/J$. The purple line indicates the nematic BKT transition between the PM and qFQ+CSL or qFQ phases. At lower temperatures, an Ising transition (red line) induced by the weak FM can be determined by the Binder ratio with $L=12,24,36,48$. The system acquires CSL features from the emergence of the local restrictions~\eqref{eq:constraintS} and~\eqref{eq:constraintQ} after a crossover (green dash-dotted line). The BKT transition disappears below $K/J\approx 0.546$, indicated by the purple line with squares, which is determined with $L=12,24,36,48,72,96$. The error bars of the Ising and BKT phase boundaries are smaller than the point sizes.
    }
    \label{fig:phasediagram_gs}
\end{figure*}
 

We introduce the (spin bilinear) quadrupole moments ${{\mathbf Q}_i=(Q^{x^2-y^2}_i,Q^{3z^2}_i,Q^{xy}_i,Q^{xz}_i,Q^{yz}_i)}$. 
For example, ${Q^{xy}_i \equiv S^x_i S^y_i + S^y_i S^x_i }$ (full definitions are given in the Supplemental Material (SM)~\cite{SM}). With this choice,  
the model in Eq.~\eqref{eq:hamiltonian}, up to a constant, is written as 
\begin{equation}\label{eq:hamiltonian_SQ}
{\mathcal H} = \sum_{\braket{ij}} \left[(J+ \frac{K}{2}) \mathbf{S}_i \cdot \mathbf{S}_j 
- \frac{K}{2} {\mathbf Q}_i \cdot {\mathbf Q}_j \right]-\sum_i  D_z (S_i^z)^2. 
\end{equation}
The spin dipolar and quadrupolar DOFs are now placed on equal footing
in terms of the ordering and fluctuations. Moreover, the single-ion anisotropy polarizes the $Q^{3z^2}$ component.

With the quadrupolar interaction, the conventional large-$S$ semiclassical treatment 
can no longer be applied. Since all spin states are connected 
by the dipole and quadrupole moments, to capture this property, we instead use a 
semiclassical SU(3) approximation. In the spirit of ``semiclassics,'' we 
consider an entanglement-free trial wave function, 
${\ket{\Psi}= \prod_i  \ket{\bm{Z}_i} }$, 
with 
${ \ket{\bm{Z}_i} = \sum_{\alpha = x,y,z} Z_i^{\alpha} \ket{\alpha} _i }$, 
where ${\ket{x} = \frac{\imath}{\sqrt{2}} {(\ket{1}-\ket{\bar{1}})}}$, ${\ket{y} = \frac{1}{\sqrt{2}} {(\ket{1}+\ket{\bar{1}})}}$,
and ${\ket{z} =-\imath \ket{0} }$ such that ${S^{\alpha} \ket{\alpha} =0}$. 
Such a basis is intrinsically designed for a quantum state that lacks dipolar order 
and is instead characterized by quadrupolar order. 
Here, ${\bm{Z}_i  = (Z_i^{x}, Z_i^{y}, Z_i^{z})^T}$ is a complex vector and is constrained by the 
normalization condition ${ \bm{Z}_i^\dagger \cdot \bm{Z}_i =1 }$. 
${\ket{\bm{Z}_i}}$ is known as the SU(3) coherent state that forms an overcomplete 
and non-orthogonal spin basis. 
All three spin states in the ${S=1}$ Hilbert space 
are now treated on equal footing, which is called the SU(3) approximation. 
Thus, this semiclassical SU(3) approximation can be thought of as a \emph{semiquantum} treatment. 
In this approximation, the \emph{semiclassical} variational energy of the system is given as 
$H_v = \braket{\Psi|\mathcal{H}|\Psi} $, with
\begin{equation}
\label{eq:semiclassical_H}
    H_v 
    = \sum_{\braket{ij}} J | {\bm{Z}_i^{\dagger}\cdot\bm{Z}_j }|^2 - (J + K)  |{\bm{Z}_i \cdot \bm{Z}_j}| ^2 
   +  \sum_{i} D_z |Z_i^z|^2, \nonumber 
\end{equation}
where the constant term has been dropped. 
The ground-state diagram, after $H_{v}$ is optimized directly, is presented in Fig.~\ref{fig:phasediagram_gs}(a).

\subsection{FQ state and its instability}%
\label{subsec:FQ}

\begin{figure}[!t]
    \centering
    \includegraphics[width=1.0\linewidth]{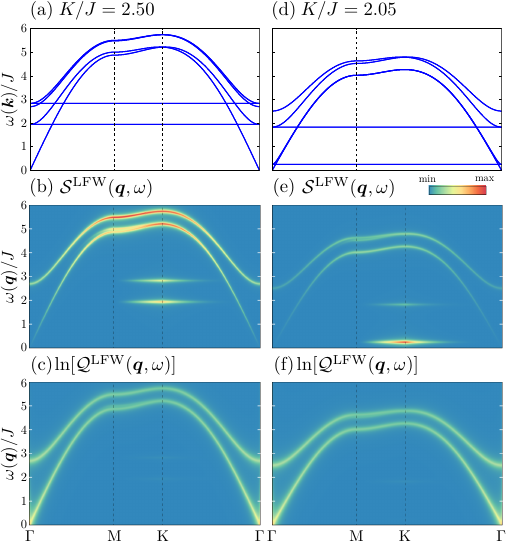}
    \caption{The evolution of spin excitation spectra in the FQ phase.
    (a) The dispersions of spin excitations, (b) dipolar dynamical SFF $\mathcal{S}^{\text{LFW}}(\bm{q},\omega)$, and (c) quadrupolar dynamical SSF $\mathcal{Q}^{\text{LFW}}(\bm{q},\omega)$. 
    The parameters $(K,D_z)/J=(2.5,1.0)$ are deep in the FQ phase. 
    (d)-(f) The equivalent results near the phase boundary with $(K,D_z)/J=(2.05,1.0)$. 
    The dynamical SSFs in the quadrupolar channel are displayed on a logarithmic scale for clarity. 
    }%
    \label{fig:LFWT}
\end{figure}

The most notable feature in the phase diagram of Fig.~\ref{fig:phasediagram_gs}(a) is that the pure FQ state, where $|\braket{\mathbf{Q}}|$ reaches its maximum, is favored when ${K/J > 2}$ regardless of the single-ion anisotropy. 
This state is time reversal invariant and has a universal $\ket{\bm{Z}_i} = \cos\phi \ket{x}_i + \sin\phi \ket{y}_i$, where $\phi$ is interaction dependent. 
The real vector $\bm{Z}$ is also known as the nematic director. 
The vanishing of the imaginary part of $\bm{Z}$ indicates the absence of the dipole moment $\braket{\mathbf{S}} = 0$. 
The presence of the easy-axis anisotropy term merely requires $\bm{Z}$ to lie on the $x$-$y$ plane at zero temperature.

The well-ordered FQ state is exact even at the quantum limit and has no classical counterpart. 
Hidden in this simple ordering, the quantum fluctuation is a bit nontrivial, especially near the phase boundary $K_c/J = 2$ due to the frustrated nature of the kagome lattice. To reveal this property, we rewrite the Cartesian basis in terms of creating three different Schwinger bosons with
\begin{equation}
    b_{i\alpha}^{\dagger}\ket{\emptyset} \equiv \ket{\alpha},\quad (\alpha=x,y,z).
\end{equation}
This representation enlarges the fundamental representation at each site, and we need to reinstate the physical Hilbert space by imposing a single-occupancy constraint,
\begin{equation}\label{eq:LFW}
    \sum_{\alpha=x,y,z}b_{i\alpha}^{\dagger}b_{i\alpha} = 1.
\end{equation} 
The Hamiltonian in Eq.~\eqref{eq:hamiltonian_SQ} can be reexpressed in terms of the Schwinger bosons,
\begin{align}
    \mathcal{H} = {} & \sum_{\braket{ij}} J b_{i\alpha}^{\dagger}b_{j\alpha}b_{j\beta}^{\dagger}b_{i\beta} - (J + K) b_{i\alpha}^{\dagger}b_{j\alpha}^{\dagger} b_{j\beta} b_{i\beta} \notag \\
    & + D_z \sum_{i} b_{iz}^{\dagger}b_{iz}, 
\end{align}
where the repeated Greek indices should be summarized over three flavors.

Without loss of generality, it is assumed that the flavor $x$ of Schwinger bosons is condensed at each site in the FQ phase via the following substitution:
\begin{equation}
    b_{i,x}^{\dagger}=b_{i,x} \approx 1 - \frac{1}{2} (b_{i,y}^{\dagger}b_{i,y}+b_{i,z}^{\dagger}b_{i,z}).
\end{equation}
After the Fourier transition, the linear flavor-wave (LFW) Hamiltonian can be expressed as
\begin{equation}\label{eq:BdG}
    H(\bm{k}) = 
    \begin{pmatrix}
        \bm{\psi}_{\bm{k}}^{\dagger} & \bm{\psi}_{-\bm{k}}
    \end{pmatrix}
    \begin{pmatrix}
        A_{\bm{k}} + C & B_{\bm{k}} \\
        B^{\dagger}_{\bm{k}} & A^{*}_{-\bm{k}} + C
    \end{pmatrix}
    \begin{pmatrix}
        \bm{\psi}_{\bm{k}} \\
        \bm{\psi}_{-\bm{k}}^{\dagger}
    \end{pmatrix},
\end{equation}
with the basis $\bm{\psi}_{\bm{k}}^{\dagger}=(b_{\bm{k}1y}^{\dagger},b_{\bm{k}2y}^{\dagger},b_{\bm{k}3y}^{\dagger},b_{\bm{k}1z}^{\dagger},b_{\bm{k}2z}^{\dagger},b_{\bm{k}3z}^{\dagger})$. 
The number index labels three sublattices. 
The technical details and the full form of $H(\bm{k})$ can be found in Ref.~\cite{SM}. 
The standard diagonalization procedure leads to six branches of LFW excitations. 
In addition to the Goldstone mode characteristic of the continuous symmetry breaking of the FQ ordering, there are two flat bands in the LFW excitation spectra. 
Specifically, one of them has the dispersion $\omega(\bm{k}) = 2\sqrt{3K(K-2J)}$, which is independent of the single-ion anisotropy $D_z$. 
With the decreasing of $K/J$, this flat band $\omega(\bm{k})$ approaches zero and vanishes at $K_c/J = 2$, as shown in Figs.~\ref{fig:LFWT}(a) and~\ref{fig:LFWT}(d), suggesting strong instability of the FQ order. 
This naturally explains the $D_z$ independence of the FQ phase boundary in Fig.~\ref{fig:phasediagram_gs}(a). 

To elucidate the impact of quantum fluctuations, we calculate the dynamical spin structure factors (SSFs) in the dipole and quadrupole channels [denoted by $\mathcal{S}^{\text{LFW}}(\bm{q},\omega)$ and $\mathcal{Q}^{\text{LFW}}(\bm{q},\omega)$, respectively]. 
The results with a logarithmic rescaling for $\mathcal{Q}^{\text{LFW}}(\bm{q},\omega)$ are displayed in the second and third rows of Fig.~\ref{fig:LFWT}. 
The dynamical SSFs on two flat bands are almost entirely contributed by the dipolar DOF. 
With the decreasing of $K/J$, the dynamical SSFs in the dipolar channel $\mathcal{S}^{\text{LFW}}(\bm{q},\omega)$ are drastically redistributed and gradually concentrate on the lowest flat band $\omega(\bm{k})$, especially near the momentum $\mathbf{K}$, as shown in Fig.~\ref{fig:LFWT}(e). 
However, such a broad peak cannot be regarded as a precursor of the conventional three-sublattice antiferromagnetic order as clarified in Ref.~\cite{SM}. 
In contrast, three coexisting phases are obtained at the semiclassical level, as described in the next section. 


\subsection{Semiclassical ground states in the coexistence regime}%
\label{subsec:coexistence_regime}

After this phase transition, the dipole moments acquire nonzero expectation values $\braket{\mathbf{S}_i}=\bra{\bm{Z}_i}\mathbf{S}_i\ket{\bm{Z}_i}$
and increase gradually with the decreasing of $K/J$. 
Three phases with distinct symmetries can be identified, and in each phase, 
the optimization of $H_v$ gives three different single-site states. 
The only restriction on optimizing $H_v$ is that each triangular plaquette 
should host all three single-site states in a given configuration. 
This restriction implies an extensive degeneracy on the kagome lattice. 
The configuration of dipole moments can be decomposed 
into two parts, as shown in the inset in Fig.~\ref{fig:phasediagram_gs}(a) for phase I, 
\begin{equation}
    \sum_{i\in\triangle}\braket{\mathbf{S}_i} = 3 \mathbf{S}_{\text{FM}} 
    + \sum_{i\in\triangle} \braket{\mathbf{S}'_i} = 3 \mathbf{S}_{\text{FM}},
    \label{eq:constraintS}
\end{equation}
where the ferromagnetic (FM) component $\mathbf{S}_{\text{FM}}$ is nonzero 
for ${D_z>0}$ and spontaneously breaks the time-reversal symmetry.
The remaining part serves as a local constraint on each triangular plaquette 
${\sum_{i\in\triangle} \braket{\mathbf{S}_i'}=0}$ with unequal length $|\braket{\mathbf{S}'_i}|$.
Here, the dipolar and quadrupolar DOFs are intricately intertwined in the semiclassical SU(3) approximation. 
Corresponding to the dipole moments, three different quadruple moments $\braket{\mathbf{Q}_i}=\bra{\bm{Z}_i}\mathbf{Q}_i\ket{\bm{Z}_i}$ are obtained. 
Similarly,
a nonzero FQ component can be separated,
\begin{equation}
    \sum_{i\in\triangle}\braket{\mathbf{Q}_i} = 3 \mathbf{Q}_{\text{FQ}} + \sum_{i\in\triangle} \braket{\mathbf{Q}'_i} = 3 \mathbf{Q}_{\text{FQ}},
    \label{eq:constraintQ}
\end{equation}
where the remaining part analogously serves as a local constraint ${\sum_{i\in\triangle} \braket{\mathbf{Q}'_i} = 0}$ on quadrupole moments.
The two local constraints for $\braket{\mathbf{S}'_i}$ and $\braket{\mathbf{Q}'_i}$ 
are the hallmarks of CSLs and are responsible for the extensive degeneracy of the ground states and the low-temperature spin correlations~\cite{annurev-conmatphys-070909-104138}. 
Phases II and III are identified with different spin configurations 
within the triangular plaquette but can be understood in the same manner~\cite{SM}. 
We thus regard the phases in the coexistence regime as a composite 
of a CSL and ferroicities for both dipolar and quadrupolar DOFs.

\section{Finite temperature phases and their properties}%
\label{sec:finite_temperature}

\subsection{Semiclassical Monte Carlo method}%
\label{subsec:sMC}

In parallel to the optimization of ground-state energy, the finite-temperature spin correlations for both dipolar and quadrupolar channels 
can be evaluated using a modified self-consistent Gaussian approximation~\cite{SM}. %
Although certain features are not well captured by this approximation due to the loss of SU(3) commutation relations, it does 
qualitatively capture some of the spin correlations. 
To better extract the finite-temperature properties of Eq.~\eqref{eq:hamiltonian}, 
we resort to the semiclassical Monte Carlo (sMC) algorithm~\cite{PhysRevB.79.214436}. 
This method is based on the variational energy $H_v$ that is extended to finite temperatures. 
The sampling in the simulation, with a standard Metropolis algorithm as the primary method, is made up of the entanglement-free direct product of quantum states 
defined previously by $\ket{\Psi}$. 
The partition function $\mathcal{Z} =  \Tr[ \exp(-\beta \mathcal{H})]$ can be approximated as 
\begin{equation}
    \mathcal{Z} \approx \int \prod_i d \bm{Z}_i e^{-\beta \bra{\Psi } \mathcal{H} \ket{\Psi}} = \int \prod_i d \bm{Z}_i e^{-\beta H_v}, 
\end{equation}
where $H_v$ is a real number.  
Now the partition function is an integration over a set of   
complex vectors $\bm{Z}_i$'s without the infamous sign problem. 
In every Metropolis update, a random site $i$ is selected, and a new vector $\bm{Z}^{\prime}_i$ is proposed near the current vector $\bm{Z}_i$ randomly. 
The energy difference $\Delta E=E_{\text{new}}-E_{\text{current}}$ is calculated between the energies of the new configuration, $E_{\text{new}}$, and of the current configuration, $E_{\text{current}}$. 
If $\Delta E<0$, the vector on site $i$ is updated to $\bm{Z}_i'$. 
Otherwise, this update is accepted with a probability $e^{-\beta \Delta E}$. 
In addition, the overall phase of each site is considered to make the random sampling $\bm{Z}^{\prime}_i$ more efficient. This step will not change the physics because all the observables we are interested in are gauge invariant.

Meanwhile, a microcanonical relaxation technique proposed in Ref.~\cite{Yamamoto2020quantum} is employed to reduce autocorrelation, which is similar to the over-relaxation technique used in the classical continuous spin model~\cite{Brown1987overrelaxed,Creutz1987overrelaxation}. 
Specifically, a local unitary transformation $e^{\imath c \mathcal{H}_i^{l}}\ket{\bm{Z}_i}$ is introduced. 
The energy of the system within the semiclassical SU(3) approximation is preserved under this transformation. 
The local operator $\mathcal{H}_i^l$ on site $i$ is defined by $\mathcal{H}_i^l = (\otimes_{j\neq i}\bra{\bm{Z}_i}) \mathcal{H} (\otimes_{j\neq i}\ket{\bm{Z}_i})$, or, more explicitly, 
\begin{equation}
    \mathcal{H}_i^l = \sum_{\braket{ij}} \left[(J+\frac{K}{2}) \mathbf{S}_i \cdot \braket{\mathbf{S}}_j 
- \frac{K}{2} \mathbf{Q}_i \cdot \braket{\mathbf{Q}}_j \right]-\sum_i  D_z (S_i^z)^2.
\end{equation}
The real number $c$ is chosen uniformly in the region $[-\pi \alpha^{-1}, \pi\alpha^{-1}]$, with $\alpha$ being the Frobenius norm of the matrix form of $\mathcal{H}_i^l$.

Additionally, parallel tempering is implemented near the Ising transition to overcome the non-ergodic problem~\cite{Hukushima1996exchange}. It offers a solution by running multiple replicas of the system at different temperatures simultaneously and exchanging configurations between them. The low-temperature replicas can more efficiently explore the configuration space by exchanging configurations with a high-temperature replica whose autocorrelation time is short. This leads to sampling improvements to statistically independent configurations and enables meaningful statistical analysis. In our simulations, the probability of exchanging replicas is in the range of about $0.3$ to $0.5$.

\begin{figure}[!t]
    \includegraphics[width=1.0\linewidth]{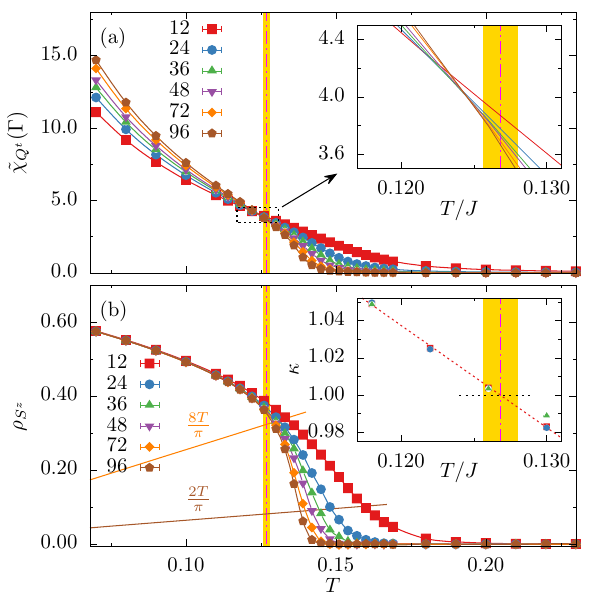}
    \caption{Scaled susceptibility $\tilde{\chi}_{Q^t}$ and stiffness $\rho_{S^z}$ for ${(K,D_z)/J=(1.0,1.0)}$.
    At the nematic BKT transition temperature $T_\text{BKT}=0.127(1)$ (pink dash-dotted line, with the yellow region acting as an error bar), (a) $\tilde{\chi}_{Q^t}$ has an intersection, and (b) $\rho_{S^z}$ has a jump. 
    The orange and brown solid lines refer to functions $f(T)=8T/\pi$ and $g(T)=2T/\pi$, respectively. 
    The inset in (b) shows solutions of RG equations and their linear fit for different pairs of system sizes.}
    \label{fig:BKT}
\end{figure}

\begin{figure}[!t]
    \includegraphics[width=\linewidth]{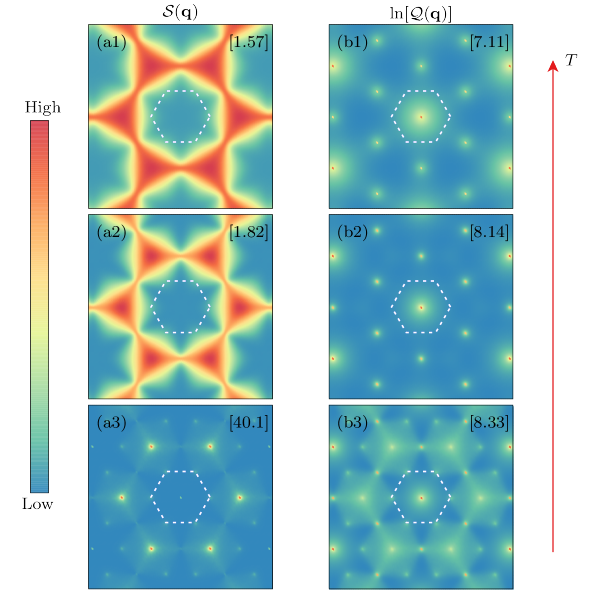}
    \caption{Dipolar and quadrupolar SSFs, $\mathcal{S}(\mathbf{q})$ and $\mathcal{Q}(\mathbf{q})$, at $(K,D_z)/J=(1.0,1.0)$ with $L=48$. The intensities are normalized with the maximum (number in the top right) in every graph. From top to bottom, the temperature decreases from ${T/J = 0.2}$ to $0.079$ to $0.016$. The system is sequentially in the PM, qFQ+CSL, and FM$_\text{CSL}$+qFQ phases. The dashed hexagon is the first Brillouin zone. The region in the momentum space corresponds to $-4\pi\leq q_{x,y}\leq 4\pi$.
    }
    \label{fig:SSF}
\end{figure}

\subsection{Finite phases and phase transitions}%
\label{subsec:fintie_temperature}

The generic finite-temperature phase diagram from the sMC simulations is depicted in Fig.~\ref{fig:phasediagram_gs}(b) with the represented value $D_z=J$. 
We utilize the specific heat, 
the fluctuation of dipolar or quadrupolar moment, 
the quadrupolar susceptibility ${\chi_{Q^t} (\mathbf{q}) = \beta\braket{|Q^t(\mathbf{q})|^2}/N}$
[where $Q^t=(Q^{x^2-y^2}, Q^{xy})$ and $N$ is the number of spins] and the spin stiffness 
$\rho_{S^z} = \{\braket{\partial^2F(\theta)/\partial\theta^2} - \beta\braket{(\partial F(\theta)/\partial\theta)^2}\}/N |_{\theta=0}$ 
[where $F(\theta)$ is the twisted free energy] to 
determine phase boundaries (see the detailed definition in the SM~\cite{SM}). 
We have simulated for other finite $D_z$'s, 
and the basic structure of the phase diagram remained unchanged.

In Fig.~\ref{fig:phasediagram_gs}(b), only one direct transition (purple line with squares) occurs
for ${K>2J}$, where the system exhibits a quasi-long-range FQ (qFQ) 
order below the transition and a true long-range spin nematic order at ${T=0}$.       
This transition is further identified as a nematic BKT transition. 
Given the failure of specific heat to determine this topological transition~\cite{Nguyen2021superfluid}, the scaled susceptibility 
$\tilde{\chi}_{Q^t}=L^{2-\eta}(\ln L)^{1/8}\chi_{Q^t}$ (where $L$ is the system size) and the stiffness $\rho_{S^z}$ are employed~\cite{Wang2021percolation}. 
For $\tilde{\chi}_{Q^t}$ in Fig.~\ref{fig:BKT}(a), the critical exponent $\eta$ is set to $1/4$ and belongs to the BKT universality class. 
At the transition, there is an intersection with different system sizes. 
The critical value of the stiffness at the transition temperature $T_c$ is $(\rho_{S^z})_c=8T_c/\pi$, 4 times larger than the conventional value as a consequence of the $\mathbb{Z}_2$ 
half vortices with the fractional vorticity of $1/2$ instead of 1 in the standard XY 
universality class~\cite{Mukerjee2006topological,Yamamoto2020quantum}. 
This fractional vorticity, rooted in the $\mathbb{R}P^2$ space of $\mathbf{Q}$, has a topological origin distinct from those in generalized XY models~\cite{Lee1985strings,Carpenter1989the,Hubscher2013stiffness,Drouin2022emergent}.
The data were further subjected to finite-size scaling using the BKT renormalization group equation $4\ln(L_2/L_1)=\int_{R_2}^{R_1}dt/[t^2(\ln t - \kappa)+t]$, where $R=\pi\rho_{S^z}/8T$ and $L_1<L_2$ are two different system sizes. The size-independent parameter $\kappa$ follows $\kappa(T) = 1+\kappa'(T_c-T)$ near $T_c$~\cite{PhysRevLett.95.237204}. The well-fitted $\kappa(T)$ curve and the critical point $\kappa(T_c)=1$ are shown in the inset of Fig.~\ref{fig:BKT}(b).
All analyses unanimously advocate the BKT transition, with the critical point indicated by pink dash-dotted lines in Fig.~\ref{fig:BKT}.

Slightly above $T_c$, a conventional crossover can be simultaneously identified in the specific heat and the susceptibility of local constraints~\cite{SM}.
It is thus attributed to the emergence of CSLs [green dash-dotted line in Fig.~\ref{fig:phasediagram_gs}(b)].
For ${0.546J \lesssim K<2J}$, while the nematic BKT transition persists, an additional transition occurs at a lower temperature with a weak ferromagnetic order [red line in Fig.~\ref{fig:phasediagram_gs}(b)].
This order is rooted in the residual FM part of Eq.~\eqref{eq:constraintS}.
Due to the time-reversal symmetry breaking, this lower-temperature transition 
is of the Ising type (see the SM~\cite{SM} for more evidence). For ${K \lesssim 0.546J}$, the nematic BKT transition 
can not be maintained given the strong dipolar frustration, 
and only the Ising transition and the CSL crossover survive, although the crossover is indistinguishable in terms of $\chi_\lambda$ (see Fig.~S8 in the SM~\cite{SM}).

To further characterize different finite-temperature phases, 
we compute the SSFs
${\mathcal{A}(\mathbf{q}) = \sum_{m,n}\braket{\mathbf{A}^m(\mathbf{q})\cdot\mathbf{A}^n(-\mathbf{q})}/N}$ ($m$ and $n$ are sublattice indexes) for both dipolar (${\mathcal{A} = \mathcal{S}}$) and quadrupolar (${\mathcal{A} = \mathcal{Q}}$) moments. 
The results for ${(K, D_z)/J = (1.0, 1.0)}$ and $L=48$ are depicted in Fig.~\ref{fig:SSF} at three different temperatures. 
For clarity, the intensities of $\mathcal{Q}(\mathbf{q})$ are taken in the logarithmic scale. 
In the featureless paramagnetic (PM) phase, broad peaks appear for both $\mathcal{S}(\mathbf{q})$ and $\mathcal{Q}(\mathbf{q})$ [Figs.~\ref{fig:SSF}(a$_1$) and (b$_1$)]. 
After a conventional crossover whose temperature is slightly higher than that of the nematic BKT type, pinch points gradually take shape at $2\mathbf{M}$ and symmetry-equivalent points in the dipolar channel, as shown in Fig.~\ref{fig:SSF}(a$_2$). 
This pattern in the SSF reveals the development of CSLs.
In the quadrupolar channel, the intensity of $\mathcal{Q}(\mathbf{q})$ in Fig.~\ref{fig:SSF}(b$_2$) is further concentrated near the $\Gamma$ point after the nematic BKT transition due to the formation of qFQ order. 
Further cooling would drive the system into a new phase via an Ising transition. 
Intensifications of $\mathcal{S}(\mathbf{q})$ and $\mathcal{Q}(\mathbf{q})$ occur at the $2\mathbf{K}$ and $\mathbf{K}$ points, while the pinch points become visible in both channels because of the enhanced effect of local constraints at the low-temperature limit. 
Within the system size we simulated, the intensity $\mathcal{S}(\mathbf{K})$ decreases with $L$ (see Fig.~S10 in the SM~\cite{SM}).
Moreover, the specific heat remains no less than $2$ (four DOFs for $\bm{Z}_i$) down to the lowest temperature, suggesting the absence of soft modes in the excitation spectrum [see Fig.~S6(a) in the SM~\cite{SM}]. 
Therefore, the thermal order-by-disorder effect, if it exists, is speculated to notably differ from the fully classical model with O(3) vector spins in which coplanar states are selected~\cite{Zhitomirsky2008octupolar,Chern2013dipolar}.
In addition, the residual FM part in Eq.~\eqref{eq:constraintS} leads to a strong peak in $\mathcal{S}(\mathbf{q})$ at the $\Gamma$ point. 
Therefore, we denote this low-temperature phase as FM$_\text{CSL}$+qFQ in Fig.~\ref{fig:phasediagram_gs}(b).

\section{Discussion}%
\label{sec:discussion}
     
Within the limit of our methods, we have obtained a rich phase diagram both at zero 
temperature and in the finite-temperature regime. The distinct patterns of spin correlations
and spin orders were carefully analyzed. 
The spin nematic type of quadrupolar order, which is sometimes 
placed into categories of hidden orders~\cite{RevModPhys.83.1301}, generates coherent spin excitations that are very 
different from conventional spin-wave excitation for conventional magnetic orders. Moreover, 
slightly akin to charge-$4e$ superconductors~\cite{Berg_2009} (but differing in the nature of the vorticity), 
the melting of the spin nematic order was found to go through a nematic type of BKT transition. 
In the ${D_z=0}$ limit on the horizontal axis in Fig.~\ref{fig:phasediagram_gs}(a), 
the symmetry of the BBQ model is restored to SU(2). 
The energy gain from the FM component $\mathbf{S}_{\text{FM}}$ vanishes in the ground states. 
The coexistence of the approximate-long-range FQ (aFQ) order and the CSL behaviors is expected. 
Moreover, the higher symmetry rules out the probability of BKT transitions from the nematic ordering. 
As a consequence, both the Ising and BKT transitions for the ${D_z = 0}$ case were found to become conventional crossovers~\cite{SM}.

The physics of the spin-1 kagome lattice antiferromagnet contains at least three major ingredients,
and they are strong geometric frustration, multipolar moments, and strong quantum fluctuations. 
In this work, we have mainly considered the interplay between the strong geometric frustration 
and the multipolar nature of the local moments. 
Owing to the choice of the method and the approximation,  
the strong quantum fluctuations were not analyzed in depth. 
When all three ingredients are taken into account, more exotic results may occur~\cite{Zhitomirsky_2010,PhysRevLett.130.116701}. 
It is well-known that, the local constraint in the classical spin liquid may turn into a matter-gauge coupled 
theory in the quantum regime. For the case of two constraints in dipolar and quadrupolar
channels in the current problem, the resulting matter-gauge couplings in the quantum regime
can have more structures and require further efforts.

\begin{acknowledgments}
    This work is supported by the MOST of China under Grants No.~2021YFA1400300 and No.~2022YFA1403902, 
    by the NSFC under Grants No.~92065203 and No.~11920101005, 
    and by the Fundamental Research Funds for the Central Universities, Peking University. 
    C.-J. H. thanks Y. Deng for valuable discussions and for his hospitality at the University of Science and Technology of China and Z. Wang for his hospitality at Zhejiang University, where part of this work was completed.
\end{acknowledgments}

\nocite{NLopt,PhysRevB.66.144407,PhysRevB.68.064411,Carrasquilla2015,Chalker1992hidden} 

\bibliography{Ref.bib}

\end{document}


\title{Supplemental Material for ``Nematic correlations and nematic Berezinskii-Kosterlitz-Thouless transition in spin-1 kagom\'{e} lattice antiferromagnets''}

\author{Chun-Jiong Huang}
\thanks{These authors contributed equally.}
\affiliation{Department of Physics and HKU-UCAS Joint Institute for Theoretical and Computational Physics at Hong Kong, 
The University of Hong Kong, Hong Kong, China} 
\affiliation{The University of Hong Kong Shenzhen Institute of Research and Innovation, Shenzhen 518057, China}
\author{Xu-Ping Yao}
\thanks{These authors contributed equally.}
\affiliation{Kavli Institute for Theoretical Sciences, University of Chinese Academy of Sciences, Beijing 100190, China}
\author{Gang v.~Chen}
\email{chenxray@pku.edu.cn}
\affiliation{International Center for Quantum Materials, School of Physics, Peking University, Beijing 100871, China}
\affiliation{Collaborative Innovation Center of Quantum Matter, Beijing 100871, China}

\date{\today}



\maketitle

\section{Optimized ground states}

The presence of the biquadratic interaction in the Hamiltonian $\mathcal{H}$ requires the finite-dimensional irreducible representation of the spin-1 moments. 
Therefore, it is more convenient to formulate the problem with the $\mathfrak{su}(3)$ Lie algebra. 
The fundamental representation of SU(3) spin operators $T^{\mu}$ (for $\mu=1,\dots,8$) can be constructed as 
\begin{subequations}
    \begin{align}
        T^7 & = + S^x, \\ 
        T^5 & = - S^y, \\
        T^2 & = + S^z, \\
        T^3 & = - [(S^x)^2 - (S^y)^2], \\
        T^8 & = - \frac{1}{\sqrt{3}}[(S^x)^2+(S^y)^2 - 2(S^z)^2], \\
        T^1 & = - [S^xS^y + S^yS^x], \\
        T^4 & = - [S^xS^z + S^zS^x], \\
        T^6 & = - [S^yS^z + S^zS^y].
    \end{align}
\end{subequations}
where $\mathbf{S}=(S^x,S^y,S^z)$ are spin-1 dipole moments that satisfy the SU(2) commutation relation $[S^{\alpha}, S^{\beta}] = \imath \epsilon_{\alpha\beta\gamma} S^{\gamma}$, where $\epsilon_{\alpha\beta\gamma}$ denotes the antisymmetric tensor. 
The later five operators with the bilinear form describe the quadrupole moments as introduced in the main text
\begin{equation}
    \mathbf{Q} = (Q^{x^2-y^2}, Q^{r^2-3z^2}, Q^{xy}, Q^{zx}, Q^{yz}) = -(T^{3}, T^{8}, T^{1}, T^{4}, T^{6}).
\end{equation}
Since they comprise the fundamental representation of $\mathfrak{su}(3)$ Lie algebra, $T^{\mu}$ are traceless 
Hermitian matrices and satisfy the orthonormality condition
\begin{equation}
    \Tr (T^{\alpha}T^{\beta}) = 2 \delta_{\alpha\beta},
\end{equation}
and the commutation relations
\begin{equation}
    [T^{\alpha}, T^{\beta}] = \imath f_{\alpha\beta\gamma} T^{\gamma}.
    \label{eq:com_rel}
\end{equation}
Note that a summation over repeated Greek indices is supposed 
hereafter unless otherwise specified. 
The factors $f_{\alpha\beta\gamma}$ are known as structure constants, 
which can be defined by the Gell-Mann matrices $\lambda_{\mu}$ as 
\begin{equation}
    f_{\eta\mu\nu} = - \frac{\imath}{2} \Tr (\lambda_{\eta}[\lambda_{\mu}, \lambda_{\nu}]).
\end{equation}
With this representation, one can prove the identity 
\begin{equation}
\mathbf{Q}_i\cdot\mathbf{Q}_j=2(\mathbf{S}_i\cdot\mathbf{S}_j)^2+\mathbf{S}_i\cdot\mathbf{S}_j-\frac{2}{3}S^2(S+1)^2,
\end{equation}
and recast the BBQ Hamiltonian $\mathcal{H}$ with the SU(3) spin operators $T^\mu$ as Equation~(2) in the main text. 

In terms of the Cartesian basis $\ket{\alpha}$ ($\alpha=x,y,z$), it is more transparent that the spin operators $T^{\mu}$ are generators of the Lie algebra associated to the SU(3) 
\begin{equation}
    \braket{\alpha|T^{\mu}|\beta} = (\lambda_{\mu})_{\alpha\beta}, 
\end{equation}
and an arbitrary spin-1 state at site $i$ can be expressed as a linear combination $\ket{\bm{Z}_i}  = Z_i^{\alpha} \ket{\alpha}_i$ where $\bm{Z}_i = (Z^x_i, Z^y_i, Z^z_i)^{\mathsf{T}}$ is a three-dimensional complex vector. 
After considering the normalization condition, a physical SU(3) coherent state 
\begin{equation}
    \ket{\bm{\Omega}} = \frac{1}{2}(1+\cos\vartheta)\exp(-\imath\varphi)\ket{1} + \frac{1}{\sqrt{2}}\ket{0}+\frac{1}{2}(1-\cos\vartheta)\exp(+\imath\varphi)\ket{\bar{1}}
\end{equation}
takes the value on a compelx projective space given by the quotient group $S^5/S^1 \cong S^5/\mathrm{U}(1) \cong \mathbb{C}P^2$.
The expectation values of SU(3) spin operators in the coherent states define the color fields
\begin{equation}\label{eq:color_field}
    n_i^{\mu} = \braket{\bm{Z}_i|T_i^{\mu}|\bm{Z}_i} = \bm{Z}_i^{\dagger}\lambda_{\mu}\bm{Z}_i.
\end{equation}
Under the semiclassical SU(3) approximation, the variational energy of the system is given by 
\begin{equation}
    H_v = \braket{\bm{Z}|\mathcal{H}|\bm{Z}} = \sum_{\braket{ij}} J |\bm{Z}_i^{\dagger}\cdot\bm{Z}_j|^2 - (J + K) |\bm{Z}_i \cdot \bm{Z}_j| ^2 + D_z \sum_{i} |Z_i^z|^2 = \sum_{\braket{ij}} \mathcal{J}^{\mu} n_i^{\mu} n_{j}^{\mu} - \frac{D_z}{\sqrt{3}} \sum_i n_i^8,
\end{equation}
where $\mathcal{J}^{\mu} = (J + K/2)(\delta_{\mu 7} + \delta_{\mu 5} + \delta_{\mu 2}) - K/2(\delta_{\mu 3} + \delta_{\mu 8} + \delta_{\mu 1} + \delta_{\mu 4} + \delta_{\mu 6})$. 
For the sake of efficiency, we first determine the number of degrees of freedom (DOF) in the $\mathbb{C}P^2$ space on each site. 
A free complex vector $\bm{Z}_i$ with three components can be generically parameterized by six variables, but the normalization condition $\bm{Z}^{\dagger}\cdot\bm{Z} = 1$ indicates a redundant global phase. 
The color field also satisfies following constraints $n^{\mu}n^{\mu} = \frac{4}{3}$ and $n^{\mu} = \frac{3}{2} d_{\mu\nu\eta} n^{\nu} n^{\eta}$, where $d_{\mu\nu\eta} = \frac{1}{4} \Tr(\lambda_{\mu}\{\lambda_{\nu}, \lambda_{\eta}\})$ are the symmetric symbols of SU(3). 
The combination of above relations lead to the Casimir identity $d_{\mu\nu\eta} n^{\mu} n^{\nu} n^{\eta} = 8/9$, which further reduces the number of DOF to four. 

At this stage, we can safely parameterize $\bm{Z}_i$ or $n_i^{\mu}$ in four independent variables at each site and conduct the optimization procedures for $H_v$ on an $L \times L$ superlattice with the periodic boundary conditions via the NLopt nonlinear optimization package~\cite{NLopt}. 
The superlattice size is taken from $L = 1$ to 6.
The optimization is repeated at least 128 times for each geometry with randomly initialized spin configurations. 
In all cases, the lowest energy is unique within the numerical error. 
Inspired by the numerical results, we further refine the ground-state configurations by taking a set of parameters with a minimum number to obtain analytic descriptions. 
Besides the FQ phase, three phases with non-zero $\braket{\mathbf{S}}$ are identified and parameterized in Tables~\ref{tab:phaseI},~\ref{tab:phaseII}, and~\ref{tab:phaseIII}, whose boundaries are presented in Fig.~1(a) of the main text.

\subsection{The pure FQ phase}

For $J/K>2$, a pure FQ phase is always favored at the level of the semiclassical SU(3) approximation. 
In the generic case where $D_z >0$, the optimized complex vectors $\bm{Z}_i$ are site-independent 
\begin{equation}
    \bm{Z}_i = [\cos\phi, \sin\phi, 0],
\end{equation}
where $\phi$ is determined by the interaction parameters $(K, D_z)/J$. 
The wavefunction of the pure FQ phase is given by the direct product $\ket{\Psi} = \prod_i \ket{\bm{Z}_i}$ where $\ket{\bm{Z}_i} = \cos\phi \ket{x}_i + \sin\phi \ket{y}_i$. 
It is easy to verify the vanishing of dipole components $\braket{S^{\alpha}_i} = 0$. 
The non-zero quadruple moments are 
\begin{equation}
    -Q_i^{x^2-y^2} = \cos2\phi, \quad -Q_i^{r^2-3z^2} = \frac{1}{\sqrt{3}}, \quad -Q_i^{xy} = \sin2\phi,
\end{equation}
Note that the expectation of $Q^{r^2-3z^2}$ is a constant due to the polarization effect from the single-ion anisotropy term $(S_i^z)^2$ when $D_z>0$.

\begin{figure}[b]
    \includegraphics[width=0.6\linewidth]{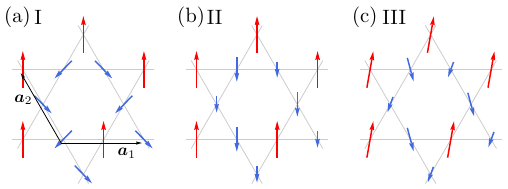}%
    \caption{The dipole spin configurations for the coexistence phases. 
    In all phases, each triangular plaquette hosts three different single-site states with distinct orders, inducing an extensive ground-state degeneracy. 
    Major dipolar spins vectors are highlighted in red.
    }%
    \label{fig:dipoleconfiguration}
\end{figure}

\subsection{The coexistence phases}

\begin{figure}[!t]
    \centering
    \includegraphics[width=0.6\linewidth]{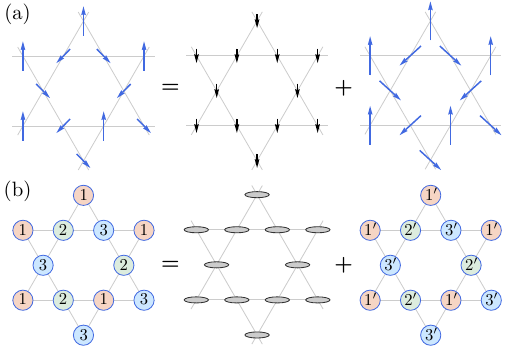}
    \caption[]{The decompositions of (a) dipole and (b) quadrupole spin components for phase I with a finite easy-axis anisotropy $D_z$. 
    There are three different vectors $\braket{\mathbf{S}_i}$ ($\braket{\mathbf{Q}_i}$) on the triangular plaquette, in which a uniform FM (FQ) part can be separated. 
    The rest part satisfies the constraint $\sum_{i\in\text{tri}}\braket{\mathbf{S}'_i}=0$ ($\sum_{i\in\text{tri}}\braket{\mathbf{Q}'_i}=0$), indicating the classical-spin-liquid physics.
    The FM part vanishes in the isotropic limit where $D_z=0$. 
    }%
    \label{fig:decomposition}
\end{figure}

The first phase, labeled by I, can be quantitatively parameterized based on the numerical optimization results.
For convenience, the $S^y$ component is fixed to be zero hereafter to eliminate the ambiguity due to the U(1) degeneracy on the $x$-$y$ plane in spin space. 
The three complex vectors on a triangular plaquette are given by 
\begin{align}
    \bm{Z}_1 = \frac{1}{\sqrt{2}}[ & (1+\imath)\cos\phi_1,-(1-\imath)\sin\phi_1,0], \\
    \bm{Z}_2 = \frac{1}{\sqrt{2}}[ & (1+\imath)\sin\theta\cos\phi_2,(1-\imath)\sin\theta\sin\phi_2, (1+\imath)\cos\theta],\\
    \bm{Z}_3 = \frac{1}{\sqrt{2}}[ & -(1+\imath)\sin\theta\cos\phi_2,-(1-\imath)\sin\theta\sin\phi_2, (1+\imath)\cos\theta],
\end{align}
where $\theta$ and $\phi_{1,2}$ are dependent on $(K, D_z)/J$. 
The explicit forms for eight spin components are listed in Table~\ref{tab:phaseI}.  
We first focus on the dipole moments. 
In the particular case where $D_z=0$, one has $\tan\phi_2 = 2\tan\phi_1$ and $\sqrt{3}\tan\theta\cos\phi_2=1$, leading to a $120^\circ$ clock configuration for dipolar components $\braket{\mathbf{S}_i}$ with a uniform but reduced length $|\mathbf{S}_i|=\sin^2 2\phi_1$ satisfying $\sum_{i\in\text{tri}} \braket{\mathbf{S}_i} = 0$. 
The presence of $D_z$ immediately introduces a finite net magnetization on each triangular plaquette to gain energy from the $(S_i^z)^2$ term. 
As a consequence, two of the three vectors $\mathbf{S}_i$ acquire a further reduced length, and their directions become more closer to the $z$-axis in spin space, albeit they are still antiparallel to the major one. 
This configuration is depicted in Fig.~\ref{fig:dipoleconfiguration}(a). 
Quantitatively, the full dipole components can be decomposed into two parts 
\begin{equation}
    \sum_{i\in\text{tri}}\braket{\mathbf{S}_i} = 3 \mathbf{S}_{\text{FM}} + \sum_{i\in\text{tri}} \braket{\mathbf{S}'_i},
\end{equation}
where the FM part is 
\begin{equation}
    \mathbf{S}_{\text{FM}} = \left( 0, 0, \frac{1}{3}\sin2\phi_1 - \frac{2}{3}\sin^2\theta\sin2\phi_2 \right).
\end{equation}
The rest part serves as a local constraint on each triangular plaquette
\begin{equation}\label{eq:constraint_S}
    \sum_{i\in\text{tri}} \braket{\mathbf{S}_i'}=0.
\end{equation}
This decomposition is intuitively depicted in Fig.~\ref{fig:decomposition}(a).  
Note that there are two types of configurations on one triangular plaquette from the complex vectors $\bm{Z}_i$. 
They are related by the time-reversal symmetry and thus characterized by opposite signs of $\mathbf{S}_{\text{FM}}$. 
Once imposed on the entire lattice, only one of them can appear in a given ground-state configuration because of the local constraint. 
Therefore, the magnetization is uniform on every triangular plaquette, resulting in a global spontaneous breaking of time-reveal symmetry and an additional $Z_2$ degeneracy. 

\begin{table*}[t]
    \begin{ruledtabular}
        \caption{The spin configuration on the triangular plaquette for dipolar and quadrupolar DOF in phase I.
        The parameters $\theta$ and $\phi_{1,2}$ are dependent on $(K, D_z)/J$. 
        The $S^y$ component has been fixed to zero for convenience. 
        }%
        \label{tab:phaseI}
        \begin{tabular}{lcccccccc}
        {} & $S^x$ & $-S^y$ & $S^z$ & $-Q^{x^2-y^2}$ & $-Q^{r^2-3z^2}$ & $-Q^{xy}$ & $-Q^{zx}$ & $-Q^{yz}$ \\
        \colrule
        1 & $0$ & $0$ & $\sin2\phi_1$ & $\cos2\phi_1$ & $\frac{1}{\sqrt{3}}$ & $0$ & $0$ & $0$ \\
        2 & $+\sin2\theta\sin\phi_2$ & $0$ & $-\sin^2\theta\sin2\phi_2$ & $\sin^2\theta\cos2\phi_2$ & $\frac{1}{\sqrt{3}}(1-3\cos^2\theta)$ & $0$ & $+\sin2\theta\cos\phi_2$ & $0$ \\
        3 & $-\sin2\theta\sin\phi_2$ & $0$ & $-\sin^2\theta\sin2\phi_2$ & $\sin^2\theta\cos2\phi_2$ & $\frac{1}{\sqrt{3}}(1-3\cos^2\theta)$ & $0$ & $-\sin2\theta\cos\phi_2$ & $0$\\
        \end{tabular}
        \end{ruledtabular}
\end{table*}

In one-to-one correspondence with the three different dipole components $\braket{\mathbf{S}_i}$, there are three different quadrupole components $\braket{\mathbf{Q}_i}$ as listed in Table~\ref{tab:phaseI}. 
An FQ part can be separated from each triangular plaquette in a similar form
\begin{equation}
    \sum_{i\in\text{tri}}\braket{\mathbf{Q}_i} = 3 \mathbf{Q}_{\text{FQ}} + \sum_{i\in\text{tri}} \braket{\mathbf{Q}'_i},
\end{equation}
where 
\begin{equation}
    \mathbf{Q}_{\text{FQ}} = \left( \frac{1}{3}\cos2\phi_1 + \frac{2}{3}\sin^2\theta\cos2\phi_2, \frac{-1}{\sqrt{3}}\cos2\theta, 0, 0, 0 \right).
\end{equation}
The rest part also serves as a local constraint but for the quadrupole moments
\begin{equation}\label{eq:constraint_Q}
    \sum_{i\in\text{tri}} \braket{\mathbf{Q}'_i} = 0.
\end{equation}
A schematic plot of this decomposition is shown in Fig.~\ref{fig:decomposition}(b) for comparison. 
In this channel, the time-reversal symmetry is always preserved. 
The two local constraints for $\braket{\mathbf{S}'}$ and $\braket{\mathbf{Q}'}$ are the hallmark of classical spin liquids and responsible for the extensive ground-state degeneracy. 
One can thus regard the ground state as a composite of a classical spin liquid and ferroicities for both dipolar and quadrupolar DOFs.  

\begin{figure}[b]
    \includegraphics[width=0.6\linewidth]{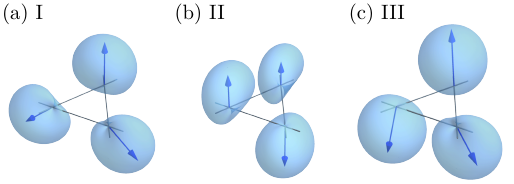}%
    \caption{The probability amplitudes $|\braket{\bm{\Omega}|\bm{Z}}|^2$ on the triangular plaquette for (a) phase I with $(K, D_z)/J = (1.0, 1.0)$, (b) phase II with $(K, D_z)/J = (1.0, 2.5)$, and (c) phase III with $(K, D_z)/J = (0.1, 2.3)$. 
    The spin configuration for dipolar DOF $\braket{\bm{S}_i}$ are indicated by blue arrows. 
    }%
    \label{fig:probability}
\end{figure}

The rest two phases can be understood in a similar manner. 
The phase II, parameterized in Table~\ref{tab:phaseII}, appears when $D_z/J$ is large enough. 
By refining the numerical results, the spin configurations can be quantitatively described by 
\begin{equation}
    \bm{Z}_i = \frac{1}{\sqrt{2}} (\cos\phi_i + \imath\cos\phi_i,-\sin\phi_i + \imath\sin\phi_i,0).
\end{equation} 
Because of the significant energy gain from the strong single-ion anisotropy, the dipolar DOF on every site are parallel/antiparallel along the $z$ direction in spin space. 
Specifically, all three dipole moments have nonequivalent expectation values as shown in Fig.~\ref{fig:dipoleconfiguration}(b) where the major dipole spins with the largest $\braket{\mathbf{S}}$ are highlighted in red. 
The rest two spins with smaller lengths are in the opposite direction of the major one. 
In the quadrupole channel, only $Q^{x^2-y^2}$ and $Q^{r^2-3z^2}$ have non-zero expectations (once the $S^y$ component is fixed to be zero), and the latter is a constant on all sublattices.

\begin{table*}[t]
    \begin{ruledtabular}
        \caption{The spin configuration on the triangular plaquette for dipolar and quadrupolar DOF in phase II.
        The parameters $\phi_{1,2,3}$ are dependent on $(K, D_z)/J$. 
        The $S^y$ component has been fixed to zero for convenience.
        }%
        \label{tab:phaseII}
        \begin{tabular}{lcccccccc}
        {} & $S^x$ & $-S^y$ & $S^z$ & $-Q^{x^2-y^2}$ & $-Q^{r^2-3z^2}$ & $-Q^{xy}$ & $-Q^{zx}$ & $-Q^{yz}$ \\
        \colrule
        1 & $0$ & $0$ & $\sin2\phi_1$ & $\cos2\phi_1$ & $\frac{1}{\sqrt{3}}$ & $0$ & $0$ & $0$ \\
        2 & $0$ & $0$ & $\sin2\phi_2$ & $\cos2\phi_2$ & $\frac{1}{\sqrt{3}}$ & $0$ & $0$ & $0$ \\
        3 & $0$ & $0$ & $\sin2\phi_3$ & $\cos2\phi_3$ & $\frac{1}{\sqrt{3}}$ & $0$ & $0$ & $0$ \\
        \end{tabular}
        \end{ruledtabular}
\end{table*}

\begin{table*}[t]
    \begin{ruledtabular}
        \caption{The spin configuration on the triangular plaquette fo dipolar and quadrupolar DOF in phase III.
        The parameters $\theta_{1,2,3}$ and $\phi_{1,2,3}$ are dependent on $(K, D_z)/J$. 
        The $S^y$ component has been fixed to zero for convenience.
        }%
        \label{tab:phaseIII}
        \begin{tabular}{lcccccccc}
        {} & $S^x$ & $-S^y$ & $S^z$ & $-Q^{x^2-y^2}$ & $-Q^{r^2-3z^2}$ & $-Q^{xy}$ & $-Q^{zx}$ & $-Q^{yz}$ \\
        \colrule
        1 & $-\sin2\theta_1\sin\phi_1$ & $0$ & $-\sin^2\theta_1\sin2\phi_1$ & $\sin^2\theta_1\cos2\phi_1$ & $\frac{1}{\sqrt{3}}(1-3\cos^2\theta_1)$ & $0$ & $-\sin2\theta_1\cos\phi_1$ & $0$\\
        2 & $-\sin2\theta_2\sin\phi_2$ & $0$ & $-\sin^2\theta_2\sin2\phi_2$ & $\sin^2\theta_2\cos2\phi_2$ & $\frac{1}{\sqrt{3}}(1-3\cos^2\theta_2)$ & $0$ & $-\sin2\theta_2\cos\phi_2$ & $0$\\
        3 & $-\sin2\theta_3\sin\phi_3$ & $0$ & $-\sin^2\theta_3\sin2\phi_3$ & $\sin^2\theta_3\cos2\phi_3$ & $\frac{1}{\sqrt{3}}(1-3\cos^2\theta_3)$ & $0$ & $-\sin2\theta_3\cos\phi_3$ & $0$\\
        \end{tabular}
        \end{ruledtabular}
\end{table*}

The last phase, III, is sandwiched between phases I and II when $K/J \lesssim 0.243$ and can be parameterized in Table~\ref{tab:phaseIII} and defined by the complex vectors 
\begin{equation}
    \bm{Z}_i = \frac{1}{\sqrt{2}} [-(1+\imath)\sin\theta_i\cos\phi_i,-(1-\imath)\sin\theta_i\sin\phi_i,
    (1+\imath)\cos\theta_i].
\end{equation}
As shown in the inset of Fig.~\ref{fig:dipoleconfiguration}(c), it has a more odd configuration with a lower symmetry for dipole moments. 
Compared with phase II, the dipolar components $\mathbf{S}_i$ in phase III can be viewed as rotations from the colinear order along a fixed direction but at different angles on different sublattices. 
Based on the same reason explained for phase I, phases II and III can also be decomposed in both dipole and quadruple channel and be regarded as a composite of a classical spin liquid and ferroicities for both dipolar and quadrupolar DOFs. 
To be more visual, Fig.~\ref{fig:probability} presents the probability amplitudes $|\braket{\bm{\Omega}|\bm{Z}}|^2$ by projecting the ground states on the SU(3) spin coherent state $\ket{\bm{\Omega}}$. 

\section{Flavor wave theory and excitations}

Since the FQ phase is well ordered, it is a good starting point to consider its instability towards the phases with $\braket{\mathbf{S}} \neq 0$ under the quantum fluctuations.  
For this purpose, we associate three Schwinger bosons $b_{i\alpha}^{\dagger}$ ($b_{i\alpha}$) with $\alpha=x,y,z$ at each site to create (annihilate) the three time-reversal invariant basis as
\begin{equation}
    b_{i\alpha}^{\dagger}\ket{\emptyset} \equiv \ket{\alpha},\quad (\alpha=x,y,z).
\end{equation}
In terms of three flavors of Schwinger bosons, the dipole and quadrupole operators are given by 
\begin{subequations}
    \begin{align}
        S_i^{\alpha} & = - \imath\varepsilon_{\alpha\beta\gamma}b_{i\beta}^{\dagger}b_{i\gamma}, \\
        Q_i^{x^2-y^2} & = - b_{ix}^{\dagger}b_{ix} + b_{iy}^{\dagger}b_{iy}, \\
        Q_i^{r^2-3z^2} & = (2 b_{iz}^{\dagger}b_{iz} - b_{ix}^{\dagger}b_{ix} - b_{iy}^{\dagger}b_{iy})/\sqrt{3}, \\
        Q_i^{xy} & = - b_{ix}^{\dagger} b_{iy} - b_{iy}^{\dagger}b_{ix}, \\
        Q_i^{yz} & = - b_{iy}^{\dagger} b_{iz} - b_{iz}^{\dagger}b_{iy}, \\
        Q_i^{zx} & = - b_{iz}^{\dagger} b_{ix} - b_{ix}^{\dagger}b_{iz}. 
    \end{align}
\end{subequations}
This representation enlarges the fundamental representation at each site and we need to reinstate the physical Hilbert space by imposing a single occupancy constraint
\begin{equation}\label{eq:LFW}
    \sum_{\alpha=x,y,z}b_{i\alpha}^{\dagger}b_{i\alpha} = 1.
\end{equation} 
The Hamiltonian in Eq.~(2) of the main text can be re-expressed in terms of the Schwinger bosons
\begin{equation}
    H_{\text{FW}} = \sum_{\braket{ij}} J b_{i\alpha}^{\dagger}b_{j\alpha}b_{j\beta}^{\dagger}b_{i\beta} - (J + K) b_{i\alpha}^{\dagger}b_{j\alpha}^{\dagger} b_{j\beta} b_{i\beta} + D_z \sum_{i} b_{iz}^{\dagger}b_{iz}, 
\end{equation}
where the repeated Greek indices should be summarized over three flavors. 

Without loss of generality, it is assumed that the flavor $x$ of Schwinger bosons is condensed at each site in the FQ phase via the following substitution
\begin{equation}
    b_{i,x}^{\dagger}=b_{i,x} =\sqrt{1-b_{i,y}^{\dagger}b_{i,y}-b_{i,z}^{\dagger}b_{i,z}} \approx 1 - \frac{1}{2} (b_{i,y}^{\dagger}b_{i,y}+b_{i,z}^{\dagger}b_{i,z}).
\end{equation}
The rest two flavors now play the roles of Holstein-Primakoff bosons and characterize the quantum fluctuations on the $y$-$z$ plane. 
After this substitution and the Fourier transformation 
\begin{equation}
    b_{i\alpha} = \frac{1}{\sqrt{N}}\sum_{\bm{k}}e^{-\imath\bm{k}\cdot\bm{r}_i}b_{\bm{k}\alpha}, \quad 
    b_{i\alpha}^{\dagger} = \frac{1}{\sqrt{N}}\sum_{\bm{k}}e^{+\imath\bm{k}\cdot\bm{r}_i}b_{\bm{k}\alpha}^{\dagger},
\end{equation}
where $N$ is the number of sites, 
The Hamiltonian $H_{\text{FW}}$ could be further recast into the bosonic Bogoliubov-de-Gennes (BdG) form in reciprocal space
\begin{equation}\label{eq:BdG}
    H(\bm{k}) = 
    \begin{pmatrix}
        \bm{\psi}_{\bm{k}}^{\dagger} & \bm{\psi}_{-\bm{k}}
    \end{pmatrix}
    \begin{pmatrix}
        A_{\bm{k}} + C & B_{\bm{k}} \\
        B^{\dagger}_{\bm{k}} & A^{*}_{-\bm{k}} + C
    \end{pmatrix}
    \begin{pmatrix}
        \bm{\psi}_{\bm{k}} \\
        \bm{\psi}_{-\bm{k}}^{\dagger}
    \end{pmatrix},
\end{equation}
with the basis $\bm{\psi}_{\bm{k}}^{\dagger}=(b_{\bm{k}1y}^{\dagger},b_{\bm{k}2y}^{\dagger},b_{\bm{k}3y}^{\dagger},b_{\bm{k}1z}^{\dagger},b_{\bm{k}2z}^{\dagger},b_{\bm{k}3z}^{\dagger})$. 
The number index labels three sublattices. 
The blocked matrix elements read $A_{\bm{k}} = J \mathbb{I}_2 \otimes \mathcal{M}_{\bm{k}}$, $B_{\bm{k}} = - (J+K)\mathbb{I}_2 \otimes \mathcal{M}_{\bm{k}}$, and $C = 2 K \mathbb{I}_6 + \frac{1}{2}D_z \diag(0,1) \otimes \mathbb{I}_3$ where
\begin{equation}
    \mathcal{M}_{\bm{k}} =
    \begin{pmatrix}
        0 & \cos\frac{\bm{k}\cdot\bm{a_1}}{2} & \cos\frac{\bm{k}\cdot(\bm{a}_1+\bm{a}_2)}{2} \\
        \cos\frac{\bm{k}\cdot\bm{a_1}}{2} & 0  & \cos\frac{\bm{k}\cdot\bm{a}_2}{2} \\
        \cos\frac{\bm{k}\cdot(\bm{a}_1+\bm{a}_2)}{2} & \cos\frac{\bm{k}\cdot\bm{a}_2}{2} & 0
    \end{pmatrix}.
\end{equation}
We have introduced the lattice vectors $\bm{a}_1 = (1,0)$ and $\bm{a}_2=(-1/2,\sqrt{3}/2)$ for the pristine kagom\'{e} lattice. 
The bosonic BdG Hamiltonian $H(\bm{k})$ can be diagonalized under the paraunitary or Krein-unitary transformation with $(\bm{\psi}_{\bm{k}}^{\dagger},\bm{\psi}_{-\bm{k}})\mathcal{U}^{\dagger}(\bm{k}) = (\bm{\beta}^{\dagger}_{\bm{k}}, \bm{\beta}_{-\bm{k}})$ where $\bm{\beta}^{\dagger}_{\bm{k}} = (\beta_{\bm{k}1y}^{\dagger},\beta_{\bm{k}2y}^{\dagger},\beta_{\bm{k}3y}^{\dagger},\beta_{\bm{k}1z}^{\dagger},\beta_{\bm{k}2z}^{\dagger},\beta_{\bm{k}3z}^{\dagger})$. 
The standard diagonalization procedure leads to 
\begin{equation}\label{eq:BdG_eigen}
    H(\bm{k}) = \sum_{\bm{k}}\sum_{m=1}^{3} \omega_{m\alpha}(\bm{k}) \beta_{\bm{k}m\alpha}^{\dagger}\beta_{\bm{k}m\alpha} + \omega_{m\alpha}(-\bm{k}) \beta_{-\bm{k}m\alpha}\beta^{\dagger}_{-\bm{k}m\alpha},
\end{equation}
where the dispersions of the linear flavor wave (LFW) excitations are given by 
\begin{widetext}
    \begin{align}
        \omega_{1}(\bm{k}) & = \left[3K(K-2J)\right]^{1/2}, \\
        \omega_{2}(\bm{k}) & = \left[(6K+D_z)(2K+D_z-4J)\right]^{1/2}, \\
        \omega_{3,4}(\bm{k}) & = \left[3K^2-(2J+K)KF(\bm{k})/2\pm(K-2J)/K/2\sqrt{3+2F(\bm{k})}\right]^{1/2}, \\
        \omega_{5,6}(\bm{k}) & = \left[D_z^2/4+D_zJ/2+2D_zK+3K^2-K(J+K/2)F(\bm{k}) \right. \notag \\
        & \hspace{10em} \left. \pm(D_zJ+2JK-K^2)/2\sqrt{3+2F(\bm{k})}\right]^{1/2}.
    \end{align}
\end{widetext}
The lattice structure factor $F(\bm{k})=\cos(\bm{k}\cdot\bm{a}_1)+\cos(\bm{k}\cdot\bm{a}_2)+\cos[\bm{k}\cdot(\bm{a}_1+\bm{a}_2)]$ has been introduced.

Because the single-ion anisotropy term is only associated with the $z$ flavor, half of the branches of spin excitations are independent of the parameter $D_z$ in the LFW theory. 
As shown in Figs.2 (a) and (d) of the main text, there is a gapless mode at the $\Gamma$ point corresponding to the Nambu-Goldstone bosons. 
This mode appears necessarily in an FQ ground state due to the spontaneous breaking of the U(1) continuous symmetry. 
In addition, there are two flat bands with eigenvalues
\begin{align}
    \omega_{1}(\bm{k}) & = \sqrt{3K(K-2J)}, \\
    \omega_{2}(\bm{k}) & = \sqrt{(6K+D_z)(2K+D_z-4J)},
\end{align}
respectively. 
The first is non-negative in the FQ phase and decreases with the biquadratic interaction $K/J$. 
It becomes zero at the phase boundary where $K/J=2$ and unphysical once $K/J < 2$, indicating the instability of the FQ phase. 
This $D_z$-independent instability is fully consistent with the semi-classical phase boundary shown in Fig.~1(a) of the main text. 

\begin{figure}[!t]
    \includegraphics[width=0.6\linewidth]{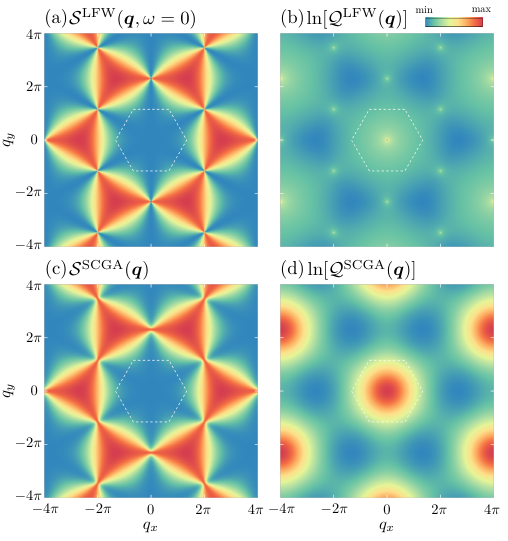}
    \caption[]{(a-b) SSFs from the LFW theory at the FQ phase boundary $(K_c,D_z)/J=(2.0,1.0)$ where $\omega_1(\bm{k})=0$. 
    The dipolar static SSF $\mathcal{S}^{\text{LFW}}(\bm{q})$ on the flat band exhibits pinch-point-like pattern. 
    The quadrupolar equal-time SSF $\mathcal{Q}^{\text{LFW}}(\bm{q})$ displayed in a logarithmic scale shows a divergent peak at $\Gamma$ point. 
    (c-d) SSFs from the SCGA theory in the phase I with $(K,D_z)=(1.0,1.0)$ at $T/J=0.1$. 
    The dashed-line hexagon indicates the first Brillouin zone.}%
    \label{fig:SSF_boundary}
\end{figure}

To elucidate the impact of quantum fluctuations, we calculate the dynamical SSFs in both dipolar and quadrupolar channels
\begin{align}
    \mathcal{S}^{\text{LFW}}(\bm{q}, \omega) & = \int_{-\infty}^{+\infty} \frac{dt}{2\pi} e^{\imath \omega t} \braket{\mathbf{S}_{\bm{q}}(t) \cdot \mathbf{S}_{-\bm{q}}(0)}, \\
    \mathcal{Q}^{\text{LFW}}(\bm{q}, \omega) & = \int_{-\infty}^{+\infty} \frac{dt}{2\pi} e^{\imath \omega t} \braket{\mathbf{Q}_{\bm{q}}(t) \cdot \mathbf{Q}_{-\bm{q}}(0)}.
\end{align}
Note that $S_i^{x}S_j^{x}$ and $Q_i^{yz}Q_j^{yz}$ do not contribute to dynamical SSFs at the LFW level because the preset nematic director $\bm{Z}$ is parallel to the $x$ axis. 
The results with a logarithmic rescale for $\mathcal{Q}^{\text{LFW}}(\bm{q},\omega)$ are displayed in the second and third rows of Fig.~2. 
Our analysis reveals that the dynamical SSFs on two flat bands are almost entirely contributed by the dipolar DOF. 
With the decreasing of $K/J$, the dynamical SSFs in the dipolar channel $\mathcal{S}^{\text{LFW}}(\bm{q},\omega)$ drastically redistribute and gradually concentrate on the lowest flat band $\omega_1(\bm{k})$, especially near the momentum $\mathbf{K}$, as shown in Fig.~2(e). 
However, such a broad peak cannot be regarded as a precursor of the conventional 3-sublattice antiferromagnetic order. 
The failure of this dipolar ordering is more transparent in the static SSF $\mathcal{S}^{\text{LFW}}(\bm{q},\omega=0)$ shown in Fig.~\ref{fig:SSF_boundary}(a). 
The static SSF for the flat band $\omega_1(\bm{k}) = 0$ on the phase boundary clearly exhibits a pinch point singularities in the dipolar channel, which is reminiscent of those observed in the Coulomb liquid phase of kagom\'{e} spin ice~\cite{PhysRevB.66.144407,PhysRevB.68.064411,Carrasquilla2015}. 
Meanwhile, the equal-time SSFs 
\begin{equation}
    \mathcal{Q}^{\text{LFW}}(\bm{q}) = \int d\omega \mathcal{Q}^{\text{LFW}}(\bm{q},\omega),
\end{equation}
in the quadrupolar channel are still dominated by the FQ correlations characterized by the divergent peak at $\bm{\Gamma}$ point. 
Although the pinch-point singularities in static SSF $\mathcal{S}^{\text{LFW}}(\bm{q},\omega=0)$ may not align with a real classical spin liquid due to the vanished expectation $\braket{\mathbf{S}}$ at the FQ phase boundary where the LFW results are already instable, these findings still light on the possibility of the coexistence of a classical spin liquid and long-rang orders from different DOF after the phase transition.

\section{Spin correlation from SCGA}

In parallel to the optimizations of ground-state energy, one can directly calculate the spin correlations at finite temperatures via the self-consistent Gaussian approximation (SCGA) under the local Hilbert-space constraint 
\begin{equation}\label{eq:local_constraint}
    |\braket{\mathbf{T}_i}|^2 = |\braket{\bm{Z}_i|\mathbf{S}_i|\bm{Z}_i}|^2 + |\braket{\bm{Z}_i|\mathbf{Q}_i|\bm{Z}_i}|^2 = \frac{4}{3},
\end{equation}
which can be verified by the definition of color fields in Eq.~\eqref{eq:color_field}. 
In this framework, the partition function of the spin-1 system in Eq.~(2) of the main text can be expressed as 
\begin{align}
    \mathcal{Z} & =\int\mathcal{D}[\mathbf{T},\lambda] e^{ -\beta H + \sum_i \lambda_i\left(\mathbf{T}_i^2 -\frac{4}{3}\right) } \approx \int\mathcal{D}[\mathbf{T},\lambda] e^{-\mathcal{S}_{\text{eff}}[\beta,\lambda_i]},
\end{align}
where an auxiliary field $\lambda_i$ has been introduced as Lagrange multipliers to enforce the local Hilbert-space constraint stated in Eq.~\eqref{eq:local_constraint} and $\beta$ is the inverse of temperature.  
At high temperatures, the fluctuation of the auxiliary field should become negligible and all lattice symmetries are expected to be restored. 
We thus perform a saddle point approximation to seek states that respect the local Hilbert-space constraint in the form of a global average $\lambda_i = - \beta \Delta(T)$ where $\Delta(T)$ is a purely real number. 
This yields an effective action
\begin{align}
    \mathcal{S}_{\text{eff}}  = {} & \beta \sum_{\bm{k}} \mathbf{T}_{\bm{k}} [\mathcal{J}(\bm{k}) + \Delta(T)\otimes\mathbb{I}_{24}] \mathbf{T}_{-\bm{k}}  - 4 \beta N  \Delta(T) \notag \\
    = {} & \beta \sum_{\bm{k}}\sum_{mn}^{3} T_{\bm{k}m}^{\mu}\left[ \mathcal{J}^{\mu}_{mn}(\bm{k}) + \Delta(T)\delta_{mn} \right]T_{-\bm{k}n}^{\mu} - 4\beta N \Delta(T),
\end{align}
where $\mathbf{T}_{\bm{k}m} = \frac{1}{\sqrt{N}} \sum_{\bm{k}} T_{\bm{r}m}e^{+\imath\bm{k}\cdot\bm{r}}$ is the Fourier transformation of spin components and $N$ is the number of unit cells. 
The indices $\bm{r}$ and $m$ are the position of unit cell and label of sublattice, respectively. 
The matrices $\mathcal{J}(\bm{k})$ are block-diagonalized in spin component space
\begin{equation}
    \mathcal{J}^{\mu}(\bm{k}) = 
    \begin{cases}
        (J+K/2) \mathsf{J}(\bm{k}) - D_z \mathbb{I}_3 \delta_{2\mu}& \mu=7,5,2\\
        -K/2 \mathsf{J}(\bm{k}) & \mu=3,1,4,6,8
    \end{cases}
\end{equation}
where $\mathsf{J}(\bm{k})$ is a $3 \times 3$ matrix 
\begin{equation}
    \mathsf{J}(\bm{k})=
    \begin{pmatrix}
        0 & \cos\frac{\bm{k}\cdot\bm{a}_1}{2} & \cos\frac{\bm{k}\cdot(\bm{a}_1+\bm{a}_2)}{2} \\
        \cos\frac{\bm{k}\cdot\bm{a}_1}{2} & 0 & \cos\frac{\bm{k}\cdot\bm{a}_2}{2} \\
        \cos\frac{\bm{k}\cdot(\bm{a}_1+\bm{a}_2)}{2} & \cos\frac{\bm{k}\cdot\bm{a}_2}{2} & 0
    \end{pmatrix}.
\end{equation}
The correlation functions between spin components $\mathbf{T}_i$ can be readily obtained 
\begin{equation}
    \braket{T_{\bm{k}m}^{\mu}T_{-\bm{k}n}^{\nu}} = \frac{1}{2\beta} \left[ \mathcal{J}(\bm{k}) + \Delta(T) \otimes \mathbb{I}_{24} \right]^{-1}_{\mu\nu,mn}.
\end{equation}
The saddle point parameter $\Delta(T)$ is determined self-consistently from the saddle point equation
\begin{equation}
    \frac{1}{N}\sum_{\bm{k}}\sum_{\mu,m}\frac{1}{2\beta}[\mathcal{J}(\bm{k}) + \Delta(T) \mathbb{I}_{24}]^{-1}_{\mu\mu,mm} = 4.
\end{equation}

In Figs.~\ref{fig:SSF_boundary}(b) and (c), we preset the static SSFs $\mathcal{S}^{\text{SCGA}}(\bm{q})$ and $\mathcal{Q}^{\text{SCGA}}(\bm{q})$ for dipole and quadrupole moments in phase I, respectively. 
Clearly, the dipolar SSF exhibits a pinch-point pattern as the one obtained from the LFW theory at the FQ boundary, although the singularities at $2\mathbf{M}$ points disappear because of the thermal broadening. 
The effect of thermal fluctuations is more significant in the quadrupolar channel. 
Compared with Fig.~\ref{fig:SSF_boundary}(b), the peak of FQ order at $\mathbf{\Gamma}$ point is melted into a hump, indicating a quasi-long-range order at finite temperatures. 

Although the features of the classical spin liquid and FQ order can be identified in SSFs obtained by the SCGA method, there is no signal of the predicted FM order in the dipolar channel even at $T/J=0.1$. 
The pinch-point pattern of the coexisting classical spin liquid is also missing in the the quadrupolar SSFs. 
In fact, the spin components $T_{i}^{\mu}$ in the SCGA method are treated as Gaussian variables that are merely subject to a relaxed global constraint. 
In contrast to the optimization in $\mathbb{C}P^2$ space, the nontrivial independency inherited from the $\mathfrak{su}(3)$ Lie algebra can not be captured by the SCGA method. 
To correctly reveal the finite-temperature properties of spin correlations in the coexistence phases, we perform the semiclassical Mento Carlo (sMC) simulations in the following sections.

\section{Observables in semiclassical Monte Carlo simulations}

To illustrate the finite-temperature properties, a collection of observables has been measured in our simulations. The following list provides the observable definitions for reference:
\begin{enumerate}
\item The specific heat $\Cv=\frac{\beta^2}{N}[\braket{E^2}-\braket{E}^2]$ where $E$ is the total energy and $N$ is the number of sites.
\item The expectation value of the operator $A(\mathbf{q})=\frac{1}{\sqrt{N}}\braket{|\mathbf{A}(\mathbf{q})|}$ at the reciprocal momentum $\mathbf{q}$, where $\mathbf{A}(\mathbf{q})=\sum_i \mathbf{A}_{i} e^{-\imath \mathbf{q}\cdot\mathbf{r}_i}$ is the Fourier transformation. Here, the operator $\mathbf{A}$ can be either the dipole operator $\mathbf{S}$, the quadrupole operator $\mathbf{Q}$ or any combination of SU(3) spin operators, the same as below.
\item The observable characterizing the local restriction
\begin{equation*}
    \lambda=\frac{3}{2N}\sum_{\triangle}\left(\sum_{i=1}^{3}\braket{\mathbf{S}_i}^2-\frac{1}{3}\left\langle\sum_{i=1}^{3}\mathbf{S}_i\right\rangle^2\right),
\end{equation*}
where $\sum_{\triangle}$ denotes the summation over all triangles in the Kagome lattice.
\item The susceptibility $\chi_A(\mathbf{q})=\frac{\beta}{N}\braket{|\mathbf{A}(\mathbf{q})|^2}$.
\item The susceptibility of the local restriction $\chi_\lambda=\frac{\beta}{N}[\braket{\lambda^2}-\braket{\lambda}^2]$.
\item The fluctuation $f_A(\mathbf{q})=\frac{1}{N}[\braket{|\mathbf{A}(\mathbf{q})|^2}-\braket{|\mathbf{A}(\mathbf{q})|}^2]$ of the operator $\mathbf{A}(\mathbf{q})$.
\item The SSF $\mathcal{A}(\bm{q})$ for the operator $\bm{A}$
\begin{equation*}
    \mathcal{A}(\bm{q})=\sum_{m,n}A^{mn}(\bm{q}),
\end{equation*}
where the matrix elements are given by 
\begin{equation*}
    {A}^{mn}(\mathbf{q})=\frac{1}{N}\braket{\mathbf{A}^m(\mathbf{q})\cdot\mathbf{A}^n(-\mathbf{q})}.
\end{equation*}
Here $m,n=1,2,3$ are the sublattice index of a kagom\'{e} lattice. The dipolar and quadruplar SSFs can be written as $\mathcal{S}(\bm{q})=\sum_{m,n}S^{mn}(\bm{q})$ and $\mathcal{Q}(\bm{q})=\sum_{m,n}Q^{mn}(\bm{q})$, respectively.
\item The spin stiffness \\
\resizebox{0.93\textwidth}{!}{$\displaystyle
\begin{aligned}
    & \rho_{S^{z}}(T) =  \frac{1}{\frac{\sqrt{3}}{2}L^2}\left.\left(\frac{\partial^2 F(\theta)}{\partial \theta^2} - \beta\left(\frac{\partial F(\theta)}{\partial \theta}\right)^2\right)\right|_{\theta=0} \\
    & = \frac{2}{\sqrt{3}L^2}\left.\left\{\left\langle\frac{\partial^{2}\left(\hat{U}_{S^{z}}(\theta) \mathcal{H} \hat{U}_{S^{z}}^{\dagger}(\theta)\right)}{\partial \theta^{2}}\right\rangle-\beta\left\langle\left(\frac{\partial\left(\hat{U}_{S^{z}}(\theta) \mathcal{H} \hat{U}_{S^{z}}^{\dagger}(\theta)\right)}{\partial \theta}\right)^{2}\right\rangle\right\}\right|_{\theta=0} \\
    & = -\frac{2}{\sqrt{3} L^{2}}\left\{\left\langle \sum_{\braket{ij}}\delta x_{ij}^{2}\left[\frac{2J+K}{2}\left(S_{i}^{x} S_{j}^{x}+S_{i}^{y} S_{j}^{y}\right)-\frac{K}{2}\left(4\left(Q_{i}^{x^{2}-y^{2}} Q_{j}^{x^{2}-y^{2}}+Q_{i}^{x y} Q_{j}^{x y}\right)+Q_{i}^{y z} Q_{j}^{y z}+Q_{i}^{x z} Q_{j}^{x z}\right)\right]\right\rangle \right. \\
    & \left. - \beta\left\langle\left( \sum_{\braket{ij}} \delta x_{ij} \left[\frac{2J+K}{2}\left( S_{i}^{x} S_{j}^{y}-S_{i}^{y} S_{j}^{x}\right)-\frac{K}{2}\left(2\left(Q_{i}^{x^{2}-y^{2}} Q_{j}^{x y}-Q_{i}^{x y} Q_{j}^{x^{2}-y^{2}}\right) - \left(Q_{i}^{y z} Q_{j}^{x z}-Q_{i}^{x z} Q_{j}^{y z}\right)\right)\right]\right)^{2}\right\rangle\right\}
\end{aligned}
$}\\
where $F(\theta)$ is the free energy under a twist angle $\theta$ around the $S^z$ axis and $\delta x_{ij}=x_i-x_j$ is the distance between the nearest neighbor spin pair $i$ and $j$ along the $x$ direction. The unitary transformation $\hat{U}_{S^{z}}(\theta)=\text{exp}(\imath\theta \sum_i x_i S_i^z)$ and $\frac{\sqrt{3}}{2}L^2$ is the Kagome lattice area.
\item The correlation function $C_{A}(r)=\braket{\mathbf{A}(r) \cdot\mathbf{A}(0)}$ where $r$ is the distance.
\item The binder ratio $R=\braket{|\mathbf{S}(\Gamma)|^4}/\braket{|\mathbf{S}(\Gamma)|^2}^2$.
\end{enumerate}
Here, $\braket{\cdots}$ means the statistical average.

\section{Extensive numerical results for the finite-temperature phases and correlations}

\begin{figure}[!t]
    \includegraphics[width=0.6\linewidth]{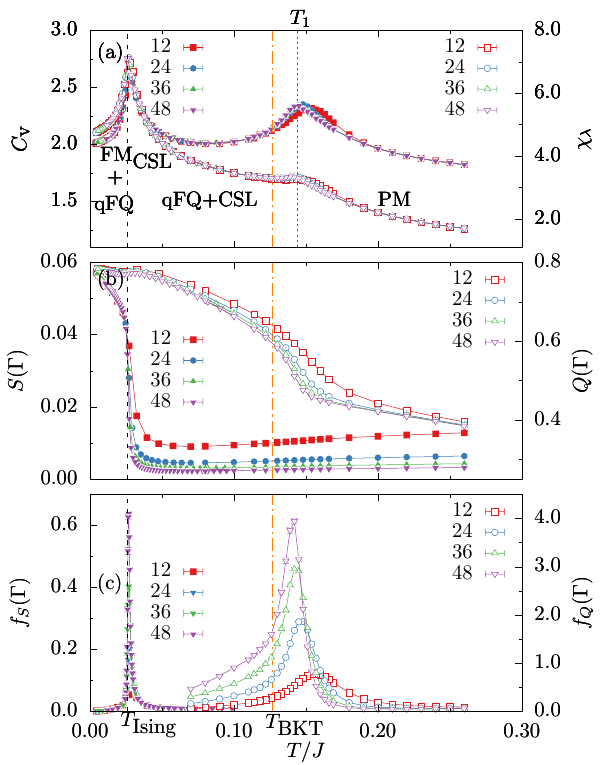}
    \caption{The evolutions of several observables over temperature at $(K,D_z)/J=(1.0,1.0)$. (a) Both the specific heat $\Cv$ and the susceptibility $\chi_\lambda$ show two peaks near temperatures $T_{\rm Ising}$ and $T_1$. (b) The quadrupole and dipole order parameters, $Q(\Gamma)$ and $S(\Gamma)$, increase rapidly near the two specific heat peaks, respectively. (c) The fluctuations, $f_S(\Gamma)$ and $f_Q(\Gamma)$. The black dashed lines, the orange dot-dashed lines and the red dotted line indicate the Ising transition, the BKT transition and the higher-temperature specific heat peak respectively. The numerical errorbars are smaller than the point size.}
    \label{fig:Dz1o00transition}
\end{figure}

To determine the finite-temperature phase diagram of Fig.~{1(b)} shown in main text, we calculate the specific heat $\Cv$, the susceptibility $\chi_\lambda$, the dipolar order parameter $S(\Gamma)$, the quadrupolar order parameter $Q(\Gamma)$, and the fluctuations $f_S(\Gamma)$ and $f_Q(\Gamma)$ in numerical simulations. Figure~\ref{fig:Dz1o00transition} shows the results for different system sizes as a function of temperature at $(K,D_z)/J=(1.0,1.0)$.
Two peaks appear in the specific heat $\Cv$ indicating large thermal fluctuations. The susceptibility $\chi_\lambda$ also shows a peak near the higher temperature $T_1/J=0.144$ demonstrating the appearance of classical spin liquid. Furthermore, near $T_1$, the quadrupolar order parameter $Q(\Gamma)$ starts to increase rapidly. Meanwhile, the fluctuation $f_Q(\Gamma)$ has a divergent peak near $T_1$.
We thus calculate the stiffness $\rho_{S^z}$ and the susceptibility $\chi_{Q^t}$ as shown in Fig.~{2} of the main text. The results indicate that there is a nematic BKT transition with an anomalous stiffness jump and it occurs at a lower temperature $T_{\rm BKT}/J=0.127(1)$. 
The inconsistency of $T_1$ and $T_{\rm BKT}$ suggests that the BKT transition is a topological phase transition and the thermal fluctuation can not reflect the transition mechanism. The same inconsistency was also reported in the two-dimensional classical XY model~\cite{Nguyen2021superfluid}.
For comparison, we plot the higher-temperature peaks' locations in the diagram (see Fig.~1(b) in the main text). As $K/J$ decreases, the difference between the two temperatures becomes smaller until $K/J\approx 0.546$ where they merge.

\begin{figure}[!t]
    \includegraphics[width=0.6\linewidth]{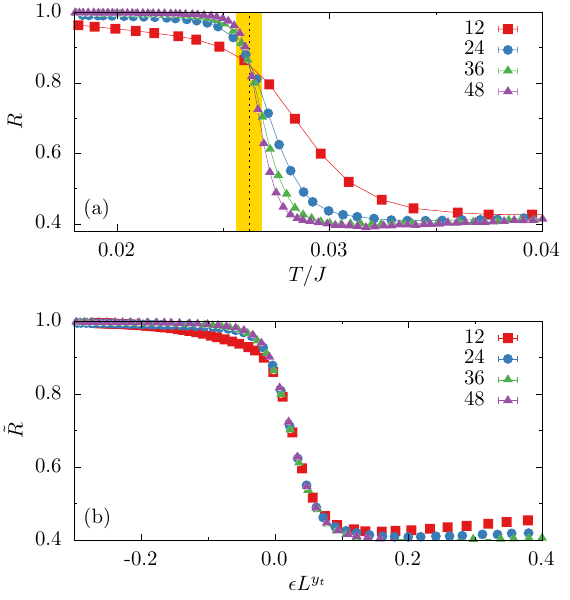}
    \caption{ 
    The binder ratio $R$ and the scaled binder ratio $\tilde{R}$ over temperature for different system sizes. (a) Near the Ising transition $T_{\rm Ising}=0.0262$ marked by a dashed line with yellow band as errorbar, binder ratio curves for different system sizes show a crossing behavior. (b) The scale binder ratio $\tilde{R}=R-b_iL^{y_i}$ versus the scaled temperature $\epsilon L^{y_t}$ with a non-universal constant $b_i=0.40$. The exponents $y_t$ and $y_i$ are set to $1$ and $-2$, respectively, according to the $2d$ Ising universality class, and $\epsilon=(T-T_{\rm Ising})$. For system sizes $L\ge 24$, the data collapse well onto a single curve. The numerical errorbars are smaller than the point size.
    }
    \label{fig:binder1}
\end{figure}

This nematic and anomalous BKT transition can be understood theoretically. 
In the presence of single-ion anisotropy, the symmetry of the BBQ model in Eq.~(1) reduces to a lower one $U(1)\times Z_2$, but the global rotation symmetry around the $S^z$ axis remains. According to the commutation relations Eq.~\eqref{eq:com_rel}, it is easy to verify the following relation
\begin{equation}
    \left[Q^{x^2-y^2}, Q^{xy}\right] = 2\imath S^z,
\end{equation}
which indicates that the quadrupolar transverse operator $Q^t=(Q^{x^2-y^2}, Q^{xy})$ has the same commutation relation to the transverse spin operator $S^t=(S^x,S^y)$ up to a factor 2. This highlights that the quadrupolar transverse operators return to their initial state after rotating by only $\pi$ around the $S^z$ axis, unlike the spin dipolar transverse operators that require a full $2\pi$ rotation. As a result, the quadrupolar transverse operators still exhibit a $U(1)/Z_2 \cong U(1)$ symmetry about the $S^z$ axis, which provides the possibility of a BKT transition. 
The additional factor of $2$ in the commutation relation originates from the nematicity of the quadruple moments and manifests itself in the nature of topological defects. 
In fact, the topological defects here are $\mathbb{Z}_2$ half vortices with the fractional vorticity $1/2$ instead of one
in the standard XY universality class. 
Therefore the critical value of the stiffness at the transition $T_c$ becomes $(\rho_{S^z})_c=8T_c/\pi$, four times than the conventional value of $2T_c/\pi$.

\begin{figure}[!t]
    \includegraphics[width=0.6\linewidth]{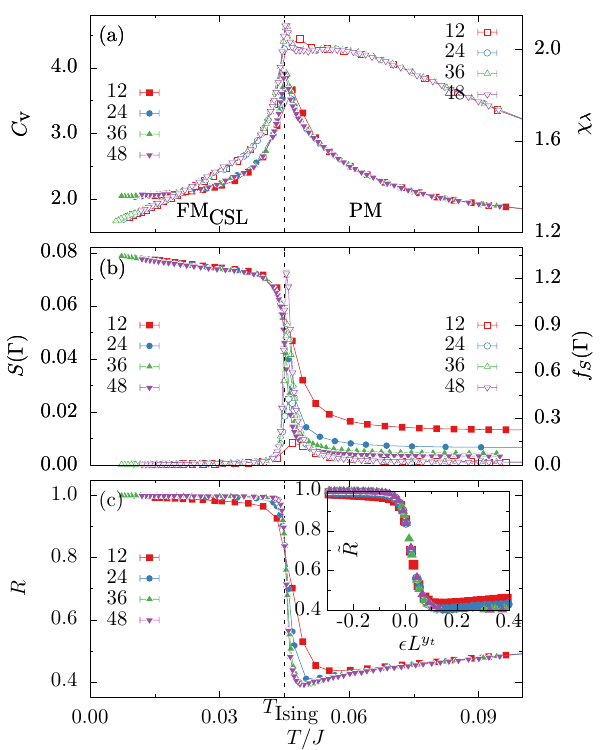}
    \caption{The evolutions of (a) the specific heat $\Cv$ and the susceptibility $\chi_\lambda$, (b) the dipole order parameter $S(\Gamma)$ and fluctuation $f_{S}(\Gamma)$, (c) the Binder ratio $R$ over temperature at $(K,D_z)/J=(0.4,1.0)$. The black dashed line indicates the Ising transition. Inset is the scale binder ratio $\tilde{R}=R-b_iL^{y_i}$ versus the scaled temperature $\epsilon L^{y_t}$ with a non-universal constant. The numerical errorbars for most data points are smaller than the point size.}
    \label{fig:Kcp0o4}
\end{figure}

As the temperature further decreases, the system enters into an FM$_{\rm CSL}$+qFQ phase (for $0.546 \lesssim K/J < 2$) or an FM$_{\rm CSL}$ phase (for $0 < K/J \lesssim 0.546$) through a $2d$ Ising transition.
To clarify this transition, the binder ratio $R$ is calculated in numerical simulations and the results are shown in Fig.~\ref{fig:binder1}(a). These curves show a crossing behavior in the vicinity of the transition $T_{\rm Ising}=0.0262(6)$.
Furthermore, Figure~\ref{fig:binder1}(b) shows curves of the scaled binder ratio $\tilde{R}$ v.s. the scaled temperature $\epsilon=(T-T_{\rm Ising})L^{y_t}$, where the critical exponents are set to $2d$ Ising universality class. The good collapse strongly advocates the transition type. Meanwhile, the susceptibility $\chi_\lambda$ has an obvious peak due to the fluctuation of the remaining ferromagnetic component $\mathbf{S}_{\rm FM}$.

\begin{figure}[!t]
    \includegraphics[width=0.6\linewidth]{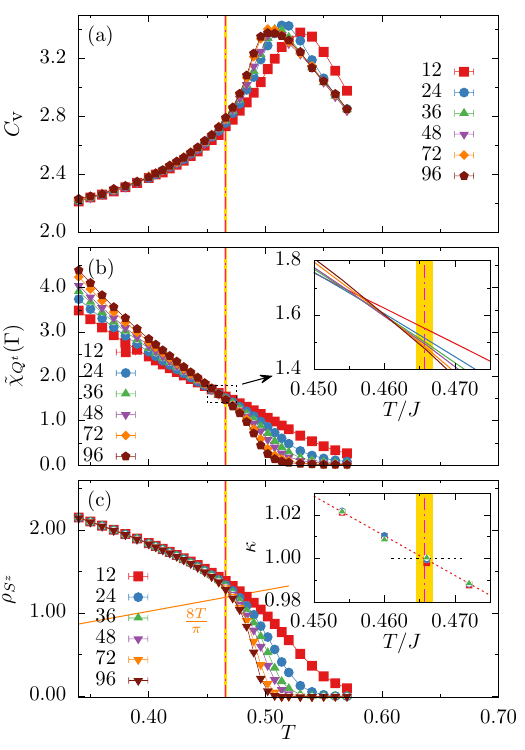}
    \caption{The evolutions of the specific heat $C_{\rm v}$, the susceptibility $\tilde{\chi}_{Q^t}$, and the stiffness $\rho_{S^z}$ over temperature at $(K,D_z)/J=(2.2,1.0)$. (a) $C_{\rm v}$ shows a round peak above the BTK transition indicating short distance correlation. (b) the scaled susceptibility $\tilde{\chi}_{Q^t}$ has an intersection near the BKT transition. (c) The stiffness $\rho_{S^z}$ has a jump near the BTK transition. Inset in (c) shows solutions of RG equations and their linear fit of the parameter $\kappa$ defined in the main text for different pairs of system sizes. The orange solid line indicates the function $f(T)=8T/\pi$ and the BKT transition is 0.466(1) indicated by an orange dash-dotted line with yellow band as errorbar. The numerical errorbars are smaller than the point size.}
    \label{fig:K2o2}
\end{figure}

We have carried out detailed numerical simulations at $(K,D_z)/J=(0.40,1.00)$ and $(2.20,1.00)$ as well.
The results of $(K,D_z)/J=(0.40,1.00)$ are shown in Fig.~\ref{fig:Kcp0o4} and only the Ising phase transition exists where the stiffness decreases as system size increases even in very low temperatures. The good collapse of the inset in Fig.~\ref{fig:Kcp0o4}(c) demonstrates that this is also a $2d$ Ising phase transition.

As mentioned in the main text, there is only a nematic BKT phase transition for $K/J>2$. In Fig.~\ref{fig:K2o2}, we show the results of $C_{\rm v}$, $\tilde{\chi}_{Q^t}(\Gamma)$ and $\rho_{S^z}$ for $(K, D_z)/J=(2.2,1.0)$. 
The nematic BKT transition can still be well located at $T_{\rm BKT}=0.466(1)$ using the BKT RG equation marked by an orange dash-dotted line with an errorbar. 

\section{Possible thermal order-by-disorder at low temperatures}%
\label{sec:coplanar}

\begin{figure}[!t]
    \includegraphics[width=0.60\linewidth]{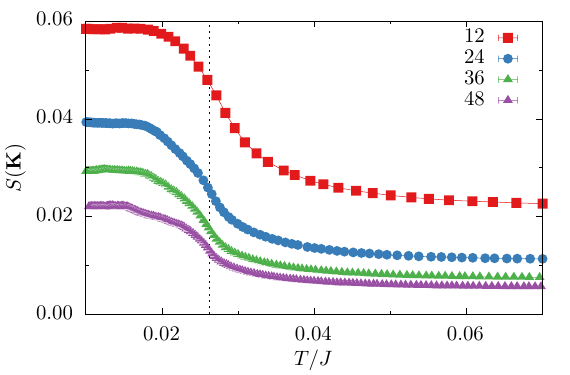}
    \caption{The evolution of the correlation functions $\mathcal{S}(\mathbf{K})$ of the dipolar moments at $(K,D_z)/J=(1.0,1.0)$. Within the simulated system sizes, $\mathcal{S}(\mathbf{K})$ monotonically decays even at low temperatures. The dashed line indicates the low-temperature specific heat peak. The numerical errorbars are smaller than the point size.}
    \label{fig:Kpoint}
\end{figure}

In the classical Kagome antiferromagnetic Heisenberg model, as temperature decreases, the system will go through three phases linked by two crossovers, i.e., the paramagnetic phase, the CSL phase, and the coplanar phase sequentially~\cite{Chalker1992hidden,Zhitomirsky2008octupolar,Chern2013dipolar}. Two significant phenomena can serve as indicators that the system has entered the coplanar phase from the CSL phase: the specific heat reduces to $11/12$ from 1 and an obvious $K$-point peak emerges in SSF. The reduced specific heat indicates the presence of zero modes which stabilizes the coplanar order and contributes only $\frac{1}{4}T$ to the internal energy less than $\frac{1}{2}T$ contributed by a quadratic mode (one degree of freedom)~\cite{Chalker1992hidden}. When the system size is small (below a hundred), the $K$-point peak in SSF shows an algebraic behavior with a large correlation length even though it approaches a finite constant at the thermodynamic limit~\cite{Chern2013dipolar}.
In our work, the enhancement of spin correlations at $\mathbf{K}$ point is clearly observed as temperature decreases and its strength monotonically decays with increasing system size up to the maximum value $L=48$ that can be simulated, as shown in Fig.~\ref{fig:Kpoint}. These results from small system sizes cannot provide sufficient evidence to support the algebraic behavior or a finite constant of $\mathcal{S}(\mathbf{K})$ at thermodynamic limit. However, Figure.~\ref{fig:Dz1o00transition}(a) shows a different specific heat behavior where the value is not less than 2 (there are four DOFs for $\bm{Z}_i$). If the system has some zero modes at low temperatures, the specific heat should be smaller than 2.
Based on the above results, we conclude that the system is still in a liquid state at low temperatures, but might be in conjunction with an approximate coplanar order.

\section{Numerical results for the SU(2) isotropic point ($D_z=0$)}

\begin{figure}[!t]
    \includegraphics[width=0.60\linewidth]{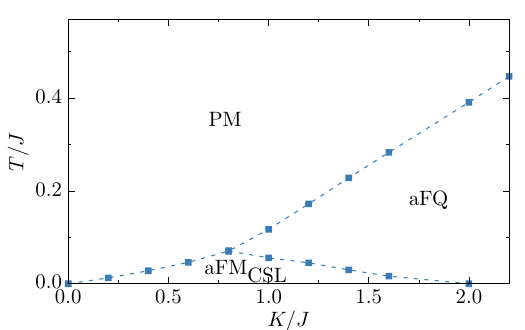}
    \caption{The finite-temperature phase diagrams at $(K,D_z)/J=(1.0,0.0)$. There are three regions: the PM region, the aFQ region, and the aFQ$_{\rm CSL}$ region. The phase boundaries are crossover and they are determined by locations of the specific heat peaks of the system size $L=36$.}
    \label{fig:Dz0o0pd}
\end{figure}

\begin{figure}[!t]
    \includegraphics[width=0.60\linewidth]{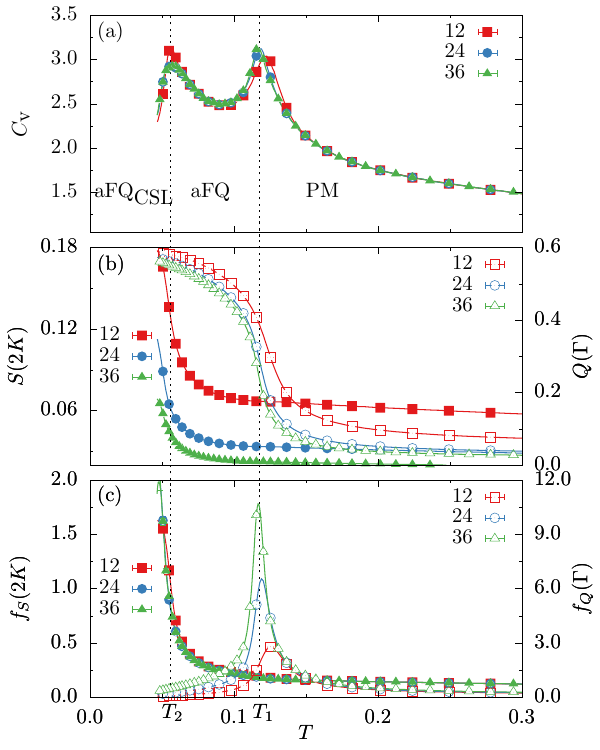}
    \caption{The evolutions of several observables over temperature at $(K,D_z)/J=(1.0,0.0)$. (a) The specific heat $\Cv$. (b) The dipole and quadrupole moments, $S(2K)$ and $Q(\Gamma)$, respectively. (c) The fluctuations, $f_S(2K)$ and $f_Q(\Gamma)$. Peaks of the specific heat at temperatures $T_1$ and $T_2$ are indicated by dashed lines.}
    \label{fig:Dz0o0}
\end{figure}

\begin{figure}[!t]
    \includegraphics[width=0.60\linewidth]{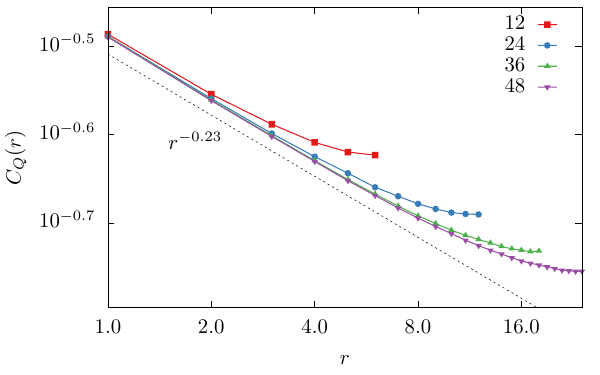}
    \caption{The correlation functions $C_Q(r)$ of the quadrupole moments at $(K,D_z)/J=(1.0,0.0)$ with the temperature $T/J=0.098$. Within the simulated system sizes, the function shows a power-law behavior before the distance reaching the half linear system size. The dash line is a guide to the eye for the power-law behavior.}
    \label{fig:CQ}
\end{figure}

In the $D_z=0$ limit, the symmetry of the model Eq.~{(1)} is restored to SU(2). We have carried out sMC simulations in this case and obtained a finite-temperature phase diagram as shown in Fig.~\ref{fig:Dz0o0pd}. To determine phase boundaries in the diagram, Figure.~\ref{fig:Dz0o0} illustrates the evolutions of the specific heat, two order parameters, and two fluctuations in the same fashion as the $D_z \neq 0$ case. 
The specific heat $\Cv$ has two peaks at a higher temperature $T_1/J=0.117$ and a lower temperature $T_2/J=0.056$. Near the lower one $T_2$, the order parameter $S(2K)$ decreases as the system size increases, and the non-divergent behavior of the fluctuation $f_S(2K)$ indicates that the transition is a crossover. At the higher temperature $T_1$, the divergence of the fluctuation $f_Q(\Gamma)$ is obvious. We further calculate the quadrupolar correlation $C_Q(r)$ at the temperature $T/J=0.098$ which is below $T_1$. It exhibits a more pronounced power-law behavior with the increase of system size, as shown in Fig.~\ref{fig:CQ}. These characteristics imply that the behavior near $T_1$ is not a second-order phase transition.
In fact, there is a possible quasi-long-range FQ order below the temperature $T_1$ according to the situation where $D_z \neq 0$.
Given the SU(2) symmetry of the model, it is worth noting that this power-law behavior may also originate from a finite-size-scaling effect due to the large correlation length which is the same as in the classical Heisenberg model in two dimensions. Therefore, we just term it an approximate-long-range FQ order (aFQ). 

\begin{figure}[!t]
    \includegraphics[width=0.60\linewidth]{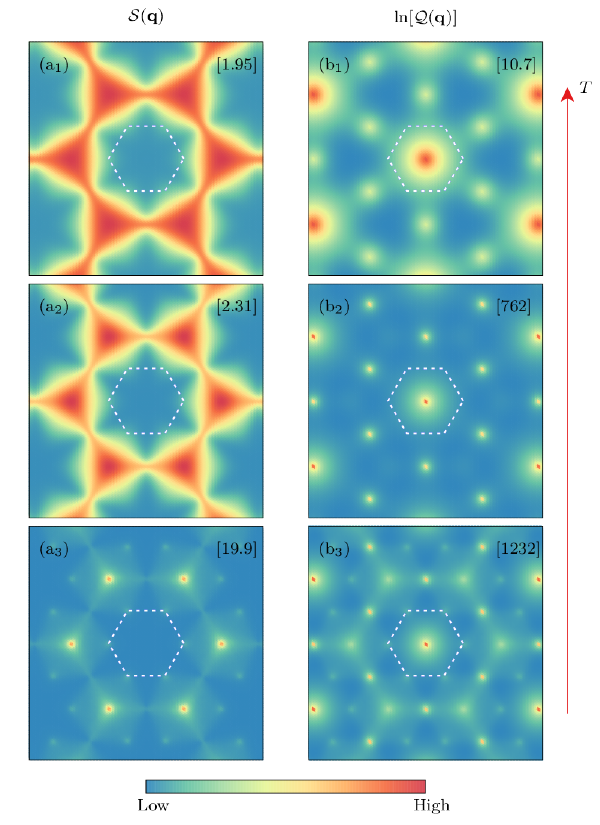}
    \caption{The dipolar and quadrupolar SSFs, $\mathcal{S}(\mathbf{q})$ and $\mathcal{Q}(\mathbf{q})$ at different temperatures when $(K,D_z)/J=(1.0,0.0)$ with the linear system size $L=36$. The intensity of SSF in each graph is normalized with its maximum presented at the top right corner. From the top to bottom, the temperature decreases and the system is sequentially in the PM phase, the aFQ phase and the aFQ$_{\rm CSL}$ phase. The temperature for each row is equal. The dash-line hexagon indicates the first Brillouin zone. Region in the momentum space corresponds to $-4\pi\leq q_{x,y}\leq 4\pi$.}
    \label{fig:Dz0o0SSF}
\end{figure}

To further characterize the system at different temperatures, we also numerically calculate the SSFs, $\mathcal{S}(\mathbf{q})$ and $\mathcal{Q}(\mathbf{q})$ as depicted in Fig.~\ref{fig:Dz0o0SSF}. 
Compared with Fig.~3 in the main text, the only qualitative difference is the disappear of $\mathbf{\Gamma}$ peak in $\mathcal{S}(\mathbf{q})$ [see Fig.~\ref{fig:Dz0o0SSF}(a$_3$)]. 
Because the energy gain from the easy-axis anisotropy term $(S_i^z)^2$ is proportional to $D_z$, the vanishing of the FM component $\mathbf{S}_{\text{FM}}$ is a natural result in the $D_z=0$ limit. 

\bibliography{Ref.bib}